\documentclass[aps, pra, 10pt, twocolumn, longbibliography]{revtex4-1}

\usepackage{textcase}
\usepackage{natbib}
\usepackage{braket}
\usepackage{amsmath}
\usepackage{graphicx}
\usepackage{hyperref}
\usepackage{amssymb}
\usepackage[utf8]{inputenc}
\usepackage{float}

\setcitestyle{square}

\makeatletter
\newcommand*\bigcdot{\mathpalette\bigcdot@{.5}}
\newcommand*\bigcdot@[2]{\mathbin{\vcenter{\hbox{\scalebox{#2}{$\m@th#1\bullet$}}}}}
\makeatother

\begin{document}
\title{Quantum Synchronisation Enabled by Dynamical Symmetries and Dissipation}

\author{J. Tindall$^{1}$, C. S\'anchez~Mu\~noz$^{1}$, B. Bu\v ca$^{1}$ and D. Jaksch$^{1,2}$}
\affiliation{$^1$Clarendon Laboratory, University of Oxford, Parks Road, Oxford OX1 3PU, United Kingdom} 

\date{\today}

\setlength{\parskip}{0pt}

\begin{abstract}
In nature, instances of synchronisation abound across a diverse range of environments. In the quantum regime, however, synchronisation is typically observed by identifying an appropriate parameter regime in a specific system. In this work we show that this need not be the case, identifying conditions which, when satisfied, guarantee that the individual constituents of a generic open quantum system will undergo completely synchronous limit cycles which are, to first order, robust to symmetry-breaking perturbations. We then describe how these conditions can be satisfied by the interplay between several elements: interactions, local dephasing and the presence of a strong dynamical symmetry - an operator which guarantees long-time non-stationary dynamics. These elements cause the formation of entanglement and off-diagonal long-range order which drive the synchronised response of the system. To illustrate these ideas we present two central examples: a chain of quadratically dephased spin-1s and the many-body charge-dephased Hubbard model. In both cases perfect phase-locking occurs throughout the system, regardless of the specific microscopic parameters or initial states. Furthermore, when these systems are perturbed, their non-linear responses elicit long-lived signatures of both phase and frequency-locking.
\end{abstract}

\maketitle

\section{Introduction} Synchronization is a fascinating and multi-disciplinary topic in modern science, focussed on understanding how a collection of individual bodies adjust their natural rhythms and phases through their interactions with each other and the environment \cite{Synch3, Synch1, Synch4, Kuramoto1, Kuramoto2}. In a striking display of cooperative behaviour, this adjustment can lead to a variety of phenomena such as the `winking' of fireflies, the behavioural synchrony of groups of strangers or the coupling of a pair of pendulums through a mutual support \cite{Fireflies, Synch2, Huygens}.
\par The study of synchronisation in quantum systems has attracted significant attention \cite{QSynch1, QSynch2, QSynch3, QSynch4, VDP2, QSynch6, QSynch65}. In this regime, synchronisation takes on a fairly broad definition due to the variety of cooperative, entangled behaviour that can occur \cite{Broad}. The formation of a Bose-Einstein condensate (BEC), for example, could be considered perfect synchronisation \cite{BECSynch1} due to the collective condensation of the atoms in the bosonic gas.  In closer analogy to classical systems, models of self-sustained quantum oscillators, such as quantum Van der Pol oscillators \cite{VDP1, VDP2, VDP3} or pairs of micromasers \cite{Micromasers}, have been shown to lock phases and reach coupled limit cycles. Quantum effects play a decisive role in either enhancing \cite{QSynch7, QSynch8} or hindering \cite{QSynch9} this synchronicity. Under the mean-field approximation, these results have been extended to larger systems of oscillators where the underlying mechanism for synchronisation is a reduction in the uncertainty in the phase distribution at the expense of the certainty in the number distribution \cite{VDP1}. 
\par Recently, there has been a focus on observing synchronisation in the limit cycles of quantum systems which have no classical analogue \cite{QSynch1, QSynch3, QSynch2}. The qutrit has been proposed as a logical candidate for this and recent work has demonstrated that it can be entrained to an external signal \cite{QSynch1, QSynch3} or phase-locked and entangled with a second spin \cite{QSynch2}. In these single or two qutrit systems, synchronisation emerges due to careful control over the Hamiltonian and dissipation parameters and is witnessed through both the phase space portrait and entanglement profile of the spins. 
\par  One of the most remarkable features of synchronisation in the classical regime, however, is that it occurs in such a diverse range of systems - with completely different sizes, structures and microscopic parameters \cite{Fireflies, Synch2, Huygens, Birds}. This diversity, in turn, leads to a rich variety of observable, complex behaviour. Hence, instead of identifying specific quantum systems and regions of parameter space where a synchronised response can be observed, we consider it pertinent to take a different route and determine, more generally, conditions which will ensure synchronisation in a quantum system. 
\par In this work we adopt this approach, identifying these conditions and uncovering a novel mechanism which guarantees synchronisation in a generic open quantum system, independent of its microscopic details. We show how these conditions can be satisfied via the interplay between several elements: interactions, local dephasing and the existence of a strong dynamical symmetry (an operator which guarantees non-stationary dynamics in the long-time limit of the system \cite{DarkHamiltonians}). The coaction of these elements underpins the formation of a structure to the long-time density matrix which ensures limit cycles describing entangled, cooperative behaviour. These cycles capture the essence of quantum synchronisation, describing oscillations where the constituents of the system are locked to a common phase and frequency whilst also featuring the off-diagonal long-range order present in states such as BECs and superconductors \cite{Condensate, Yang1}. Furthermore, we prove that this mechanism for synchronisation is, to first order, completely robust to the presence of symmetry-breaking perturbations. 
\par We then present several physical examples, which have no classical analogue, where this phenomenon arises -  a chain of interacting spin 1s and the many-body charge-dephased Hubbard model. These systems exhibit perfect distance-invariant phase synchronisation for a wide range of parameters and initial states. Moreover, in these examples, we are able to identify analytical expressions for the long-time density matrix - which is typically an unfeasible task in strongly-correlated many-body systems. 
Finally, we peturb these systems away from the dynamical symmetry regime where the non-linear response facilitates the observation of strong, exceptionally long-lived signatures of both phase and frequency locking.
\par Quantum synchronisation is sometimes viewed in terms of a locking in phase space of self-sustained oscillators, measured through the Husimi-Q or Wigner phase space distributions \cite{QSynch1, QSynch2, VDP3}. In the work and examples in this manuscript we have, instead, opted to focus on the explicit limit cycles of the bodies in the system and plot them alongside the various quantum synchronisation measures we use (such as the entanglement or off-diagonal coherences). This is in order to reflect our intuition of quantum synchronisation as an intrinsically rhythmic process underpinned by quantum properties not available in classical systems. Our definition exposes the presence of quantum properties in the synchronised states which are less easy to see in, for example, the Husimi-Q distribution. 

\section{Synchronisation in generic quantum systems}
\subsection{Strong Dynamical Symmetries} 
Firstly, we introduce the concept of a strong dynamical symmetry by providing a brief summary of the work in \cite{DarkHamiltonians}, we restrict ourselves to Markovian dynamics for simplicity. Consider the time evolution of the density matrix of an open quantum system via the Lindblad equation (here, and in the remainder of this work, we set $\hbar =1$)
\begin{align}
\frac{\partial \rho}{\partial t} = \mathcal{L}\rho &= -i[H, \rho] + \sum_{j}\gamma_{j}\big(L_{j}\rho L_{j}^{\dagger} - \frac{1}{2}\{L^{\dagger}_{j}L_{j}, \rho\}\big) \ \notag \\
&= - i [H, \rho] + D[\rho],
\label{Eq:Master Equation}
\end{align}
where $H$ is the Hamiltonian of the system and $\{L_{j}\}$ are a set of `jump' operators which model the interaction between the system and the environment with associated coupling strengths $\{\gamma_{j}\}$. The jump operators are used to form the dissipator $D[\rho]$ which competes with the coherent evolution due to the Hamiltonian $H$. We denote the Liouvillian superoperator with $\mathcal{L}$ and the steady state(s) as $\rho_{ss}$, which satisfy $\mathcal{L}\rho_{ss} = 0$. 
\par If we can identify an operator $A$ which satisfies
\begin{equation}
[H, A] = \omega A, \quad [L_{j}, A] = [L^{\dagger}_{j}, A] = 0 \quad \forall j, \ \omega \in \mathbb{R}
\label{Eq:Conditions}
\end{equation}
then we say that the system posseses a `strong dynamical symmetry'. The relation $[H, A] = \omega A$ describes the presence of a dynamical symmetry operator. We then refer to this as a strong dynamical symmetry operator because it commutes with all the jump operators and their conjugates \cite{Prosen}. 
\par It is then straightforward to prove from these relations that there exists a series of eigenmodes of $\mathcal{L}$ of the form
\begin{equation}
\rho_{n,m} \propto (A)^{n}\rho_{ss}(A^{\dagger})^{m}, \quad \mathcal{L}\rho_{n,m} = i\omega(m-n)\rho_{n,m},
\label{Eq: Imaginary Modes}
\end{equation}
where the corresponding imaginary eigenvalues indicate the presence of non-stationary dynamics in the long-time limit of the system. The operator $A$ acts as a raising/lowering operator, generating a ladder of equidistant mixed states within the kernel of the Liouvillian. These results extend beyond that of a decoherence free subspace \cite{DFS1, DFS2} as the imaginary modes are, in general, mixed and cannot be written as a convex superposition of pure states $\ket{\phi}\bra{\phi}$ which are immune to the dissipation $L_{j}\ket{\phi} = 0 \ \forall j$. 
\subsection{Quantum Synchronisation via Dynamical Symmetries, Interactions and Dephasing}
\label{Sec: Quantum Synchronisation via Dynamical Symmetries}
\par We now show how a generic open quantum system can provide a natural environment for observing quantum synchronisation. Consider an open quantum system where the Hilbert space is constructed from a series of $N$ identical, local spaces or `bodies' $\mathcal{H} = \otimes_{j}\mathcal{H}_{j}$.
\par We now imagine the system has a strong dynamical symmetry operator satisfying Eq. (\ref{Eq:Conditions}) and assume the imaginary modes from Eq. (\ref{Eq: Imaginary Modes}) form a complete basis for the long-time density matrix of the system. Conseqently, we can write this state as
\begin{equation}
\lim_{t \rightarrow \infty} \rho(t) = \rho_{\infty}(t) = \sum_{n,m;
\ n \geq m}\left(C_{n,m}e^{i \omega (m-n) t}\rho_{n,m} + {\rm h.c.}\right),
\label{Eq: DMEvolution}
\end{equation}  
where the $C_{n,m}$ are a set of real coefficients associated with the overlap between the initial state and the $\rho_{n,m}$. Now, consider the expectation value of some $M$-point observable $X = \prod_{j \in B}X_{j}$, where $B = \{a,b,c, ...\}$ is a set of $M$ local spaces containing no duplicates and $X_{j}$ is some hermitian local operator acting on site $j$. It follows from Eq. (\ref{Eq: DMEvolution}) that
\begin{equation}
\lim_{t \rightarrow \infty} \langle X \rangle (t) = \sum_{n,m; \ n > m} D_{n,m}\cos \left(\omega (m-n)t \right) + {\rm const.},
\label{Eq: Evolution}
\end{equation}
with $D_{n,m} = 2{\rm Tr}(X\rho_{n,m})C_{n,m}$. Provided that $D_{n,m} \neq 0$ for at least one $n,m$ where $|n-m| = 0$ then Eq. (\ref{Eq: Evolution}) describes coherent, non-decaying limit cycles in the associated observable. The equidistance of the imaginary eigenspectrum is crucial and ensures the frequencies involved are commensurate and do not destructively interfere with each other. These limit cycles, along with the well-defined, coherent phase-evolution described in Eq. (\ref{Eq: DMEvolution}) are some of the hallmark features of temporal synchronisation.
\par Importantly, to have full synchronisation we need each of the bodies to undergo the same coherent phase evolution. Whilst the existence of a strong dynamical symmetry ensures non-stationarity, it does not mean that the bodies will lock together in phase space and undergo identical limit cycles. The fundamental requirements for this to happen are that the steady state and strong dynamical symmetry operator(s) are translationally invariant and, as assumed earlier, form a complete basis for the long-time density matrix $\rho_{\infty}(t)$ of the system. If these requirements are met then, through Eq. (\ref{Eq: Imaginary Modes}), the $\rho_{n,m}$ and thus $\rho_{\infty}(t)$ will inherit the translational symmetry of these operators and the limit cycles described in Eq. (\ref{Eq: Evolution}) will be independent of the specific bodies in the set $B = \{a,b,c, ...\}$ (only the cardinality of the set matters). Consequently, the system will be perfectly synchronised as all the bodies in the system will be locked to the same frequency and phase - independent of the specific value of any microscopic parameters. 
\par Recent work has shown that the interplay between interactions and local, homogeneous dephasing in an open quantum system can `wash' out any geometry associated with the system: creating steady states with off-diagonal long-range order and ensuring they, along with any strong dynamical symmetry operators, are completely translationally symmetric (see \cite{DarkHamiltonians, Tindall}). These states and operators form a complete basis for the long-time density matrix of the system. Hence, following the discussion in the previous paragraph, we identify interactions between the bodies in our system as well as local, homogeneous dephasing \footnote{local and homogeneous in the sense the $L_{j}$ in Eq. (\ref{Eq:Master Equation}) are purely local and each site experiences the same jump operators and dissipation rates.
} as elements which, when combined with the existence of a strong dynamical symmetry, can ensure the system reaches a completely quantum synchronised state: i.e. with locked limit cycles underpinned by intrinsically quantum properties such as entanglement and off-diagonal long-range order. We will illuminate these ideas with a pair of examples in Sec. 3 and explicitly show these fully synchronised cycles alongside their intrinsically quantum behaviour.
\par We anticipate that for synchronisation to occur in our framework via homogeneous, local dephasing the local Hilbert space dimension should satisfy ${\rm Dim}(H_{j})>2$. This is because there must be local coherences available in the long-time limit where a valid phase relationship can be established and the system can undergo long-time oscillations. Any non-trivial local dephasing in an array of 2-level systems will destroy the available coherences and prevent the qubits from undergoing a valid limit-cycle, a pre-requisite for synchronisation. This argument does not apply to arrays of 2-level systems under more general non-local dissipation.
\par In this work we consider synchronisation under the Markov approximation and so are limited to weakly interacting systems. However, in our examples synchronisation occurs for any finite interaction strength – its amplitude only sets the timescale on which a synchronised state is reached. Hence, even if the interaction strength is small, the system will eventually reach a synchronised state (in experimental setups care should also be taken that the timescale on which synchronisation occurs is shorter than the coherence time of the system). Moreover, we emphasize that highly controllable quantum systems such as lattices of ultracold atoms immersed in a Bose-Einstein Condensate \cite{DarkHamiltonians, Tindall}, can be engineered to accurately implement dynamics described by local master equations. 
\par 

\subsection{Perturbations away from the Dynamical Symmetry Regime}
\label{Sec: Perturbations away from the Dynamical Symmetry Regime}
We now show how the synchronisation discussed in the previous section is robust to perturbations away from the dynamical symmetry regime. Typically, we can imagine that the relation $[H, A] = \omega A$ from Eq. (\ref{Eq:Conditions}) arises due to some homogeneous field $F = \omega\sum_{j}f_{j}$ in the Hamiltonian for which $A$ is a raising/ lowering operator. 
The synchronisation described in the previous section is then a consequence of perfect phase locking of the individual constituents of the systems, frequency locking will occur because the individual bodies $j$ share the same natural frequency $\omega$.
\par If the field is not homogeneous, i.e. $F = \sum_{j}\omega_{j} f_{j}$, then the Hamiltonian $H$ can be split into two terms. The first contains a homogeneous term $\sum_{j} \bar{\omega}_{j} f_{j}$ and all other non-field terms, the second contains only the inhomogeneous part $\sum_{j} \delta_{j} f_{j}$. We have parametrised $\omega_{j}$ as $\omega_{j} = \bar{\omega}_{j} + \delta_{j}$ where $\bar{\omega}_{j}$ is the average of the set $\{\omega_{j}\}$. We then, correspondingly, split the Liouvillian into two parts and scale by $1 / \bar{\omega}_{j}$:
\begin{eqnarray}
&\mathcal{L} = \mathcal{L}^{(0)} + \epsilon \mathcal{L}^{(1)}, \notag \\ &\mathcal{L}^{(0)} = -\frac{i}{\bar{\omega}_{j}}[H - \sum_{j}\delta_{j} f_{j}, \bigcdot] + \frac{1}{\bar{\omega}_{j}}D[\bigcdot], \notag \\ &\mathcal{L}^{(1)} = -i \left[ \sum_{j}\frac{\delta_{j}}{\bar{\delta}_{j}} f_{j}, \bigcdot \right],
\label{Eq:ModLiouvillian}
\end{eqnarray}
where $\epsilon = \bar{\delta}_{j} / \bar{\omega}_{j}$ is the, small, perturbation parameter and $\bar{\delta}_{j}$ is the average over the set of detunings $\{\delta_{j}\}$. We then assume that the eigenvectors and eigenvalues of this new Liouvillian are perturbations on those for $\mathcal{L}^{(0)}$
\begin{align}
\rho &= \rho^{(0)} + \epsilon \rho^{(1)} + \epsilon^{2}\rho^{(2)} + \hdots, \notag \\
\lambda &= \lambda^{(0)} + \epsilon \lambda^{(1)} + \epsilon^{2}\lambda^{(2)} + \hdots.
\label{Eq: Perturb}
\end{align}
It then follows (see Supplemental Material, SM) that the first order eigenvalue shift is 
$\lambda^{(1)} = {\rm Tr}\left[ \left( \rho^{(0)} \right)^{\dagger} \mathcal{L}^{(1)} \rho^{(0)} \right]$,
which is purely imaginary as it can be rearranged to be the trace of a skew-hermitian matrix. As a result, to first order, the eigenvalues of the eigenmodes in Eq. (\ref{Eq: Imaginary Modes}) remain imaginary and thus there is no decay in the system's long-time dynamics when perturbed away from the dynamical symmetry regime.
\par Moreover, if the imaginary eigenmodes are unchanged under a swap between two bodies $j$ and $l$ then we have that $\lambda^{(1)} = 0$ as ${\rm Tr}\left[\left( \rho^{(0)} \right)^{\dagger} f_{j} \rho^{(0)}\right]$ is independent of $j$. As discussed in Sec. \ref{Sec: Quantum Synchronisation via Dynamical Symmetries} this symmetry in the imaginary modes is seen for local, translationally-invariant dephasing in an interacting system \cite{DarkHamiltonians, Tindall}.
Hence, the system will undergo completely non-linear response to peturbations away from the dynamical symmetry regime. To first order, the solutions in Eq. (\ref{Eq: Imaginary Modes}) are still eigenmodes and so will decay with a rate that scales at least quadratically with the perturbation parameter $\epsilon$. Hence they correspond to `slow' modes which will, in general, decay much slower than rest of the eigenmodes of the Liouvillian. We can therefore expect to transiently observe the corresponding synchronised features these eigenmodes possess, with a lifetime that scales at least quadratically with the perturbation parameter $\epsilon$. The individual constituents will be locked in both phase and frequency, despite having different natural frequencies. 
\par This appearance of synchronisation due to the formation of `slow' decay modes in the Liouvillian which lock the system to specific frequencies is consistent with the mechanism for transient synchronisation discussed in \cite{Broad} and observed in \cite{PhysRevA.95.043807}.

\section{Examples}
\label{Sec:Examples}
\par We have shown how, in a generic interacting open quantum system, the combination of interactions, local dephasing and a strong dynamical symmetry can underpin a coherent, distance-invariant, synchronised structure to the long-time density matrix. Furthermore, the system is robust to perturbations away from the dynamical symmetry regime. In order to elucidate these results we present a pair of paradigmatic examples where they can be observed.
\subsection{Synchronisation in a chain of Spin-1s}
For our first example, we take a system formed from a series of spin-1s or qutrits. The local basis for each spin-1 is spanned by the three states $\{\ket{\downarrow}, \ket{0}, \ket{\uparrow}\}$. The key operators are $S^{+}_{j}$, $S^{-}_{j}$ and $S^{z}_{j}$ which are, respectively, the spin-1 raising, lowering and magnetisation operators for spin $j$. The $x$ and $y$ components of the spin-1 operator can be formed from the raising and lowering operators: $S_{j}^{x} = (1/2)(S_{j}^{+} + S_{j}^{-})$, $S_{j}^{y} = (i/2)(S_{j}^{-} - S_{j}^{+})$. By dropping the local subscript we denote the total of an operator, e.g. $S^{z} = \sum_{j}S^{z}_{j}$.

\begin{figure}[t]
\includegraphics[width = \columnwidth]{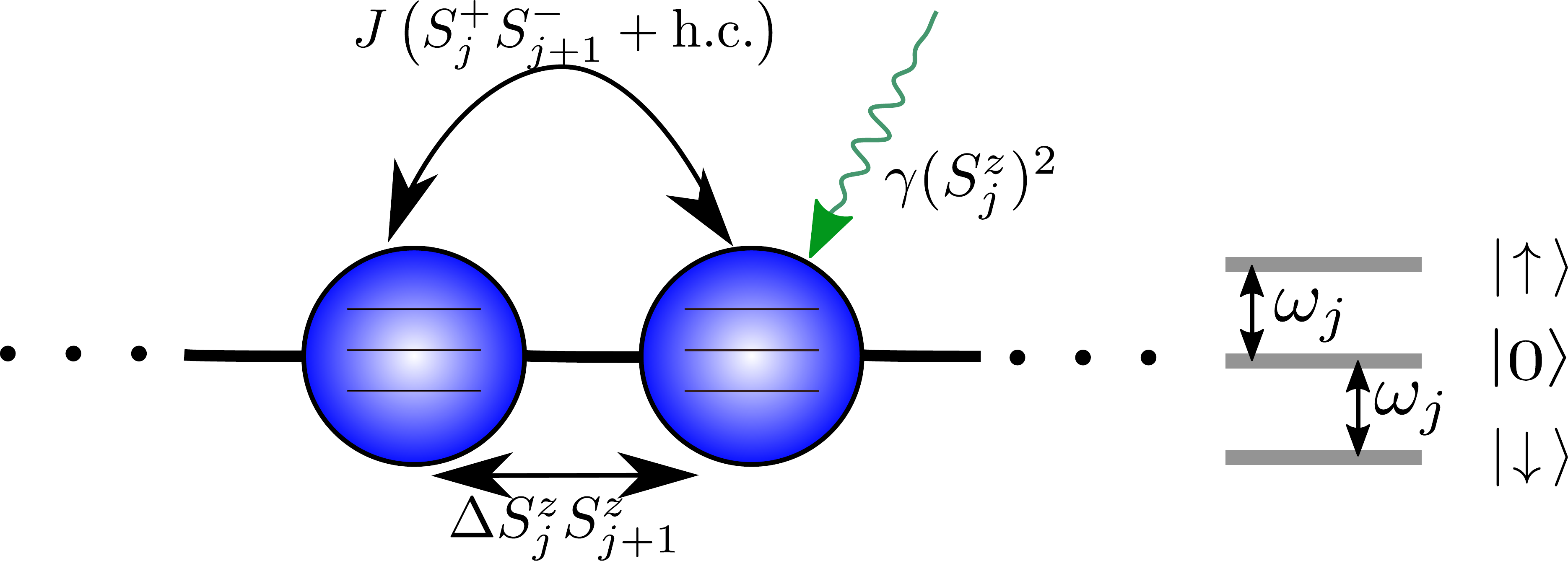}
\caption{Series of spin-$1$s in a chain geometry. The system is governed by the Hamiltonian in Eq. (\ref{Eq:Hamiltonian}) with on-site dephasing of the form $L_{j} = (S^{z}_{j})^{2}$. The resulting dynamics is described by the master Equation in Eq. (\ref{Eq:Master Equation Specific}).}
\label{Fig:Network}
\end{figure}

We take a spin-1 anisotropic Heisenberg model in a chain geometry \cite{Haldane} (see Fig. \ref{Fig:Network})
\begin{equation}
H = \sum_{j = 1}^{N} \omega_{j} S^{z}_{j} + \sum_{j = 1}^{N-1}J\big(S^{+}_{j}S^{-}_{j+1} + S^{-}_{j}S^{+}_{j+1}\big) + \Delta S^{z}_{j}S^{z}_{j+1},
\label{Eq:Hamiltonian}
\end{equation}
where the spins each have natural frequency $\omega_{j}$ and nearest-neighbour coupling strengths $J$ and $\Delta \neq 0$. The system is then immersed in a bath which induces local, quadratic dephasing in a spin-agnostic manner. The ensuing dynamics is modelled via the equation 
\begin{equation}
\frac{\partial \rho}{\partial t} = \mathcal{L}\rho = -i[H, \rho] + \gamma \sum_{j = 1}^{N}(S^{z}_{j})^{2}\rho (S^{z}_{j})^{2} - \frac{1}{2}\{(S^{z}_{j})^{4}, \rho\},
\label{Eq:Master Equation Specific}
\end{equation}
which is the master equation in Eq. (\ref{Eq:Master Equation}) with jump operators $L_{j} = (S^{z}_{j})^{2}$ applied to each spin at a rate $\gamma \ \forall j$. 

Initially, in order to derive an analytical solution to the long-time dynamics, we focus on the `frequency-matched' case $\omega_{j} = \omega, \ \forall j$. In the SM we prove that the $2N + 2$ steady states alway take the form
\begin{equation}
\rho_{ss} = \sum_{m = -N}^{N}\lambda_{m}\bigg(\sum_{i}\ket{m_{i}}\bra{m_{i}}\bigg) + \lambda_{0}' \sum_{i }\ket{0_{i}}\bra{0'_{i}},
\label{Eq:Steady State}
\end{equation}
where $\ket{m_{i}}$ is an eigenvector of $S^{z}$ with eigenvalue $m$: $S^{z}\ket{m_{i}} = m\ket{m_{i}}$, and $i$ indexes the possible eigenvectors for each $m$. We have also defined $\ket{-m'_{i}} = {\rm SF}\ket{m_{i}}$, where ${\rm SF} = \otimes_{j=1}^{N}\big(\ket{\uparrow}\bra{\downarrow} + \ket{\downarrow}\bra{\uparrow} + \ket{0}\bra{0}\big)$ is the spin-flip operator. For example if $\ket{2_{1}} = \ket{0 \uparrow \uparrow}$ then $\ket{-2_{1}'} = \ket{0 \downarrow \downarrow}$, or if $\ket{0_{2}} = \ket{0 \uparrow \downarrow}$ then $\ket{0_{2}'} = \ket{0 \downarrow \uparrow}$. In order for ${\rm Tr}(\rho_{ss}) = 1$ the elements $\{\lambda_{m}\}$ and $\lambda_{0}'$ must satisfy the equation
\begin{equation}
\lambda_{0}' + \sum_{m = -N}^{N}\lambda_{m}\sum_{s = 0}^{N}{N \choose s}{N - s \choose (N- s + m)/2} = 1,
\end{equation}
where the terms in the second summation are skipped if $(N- s + m)/2$ is not an integer.

\begin{figure}[t]
\includegraphics[width = \columnwidth]{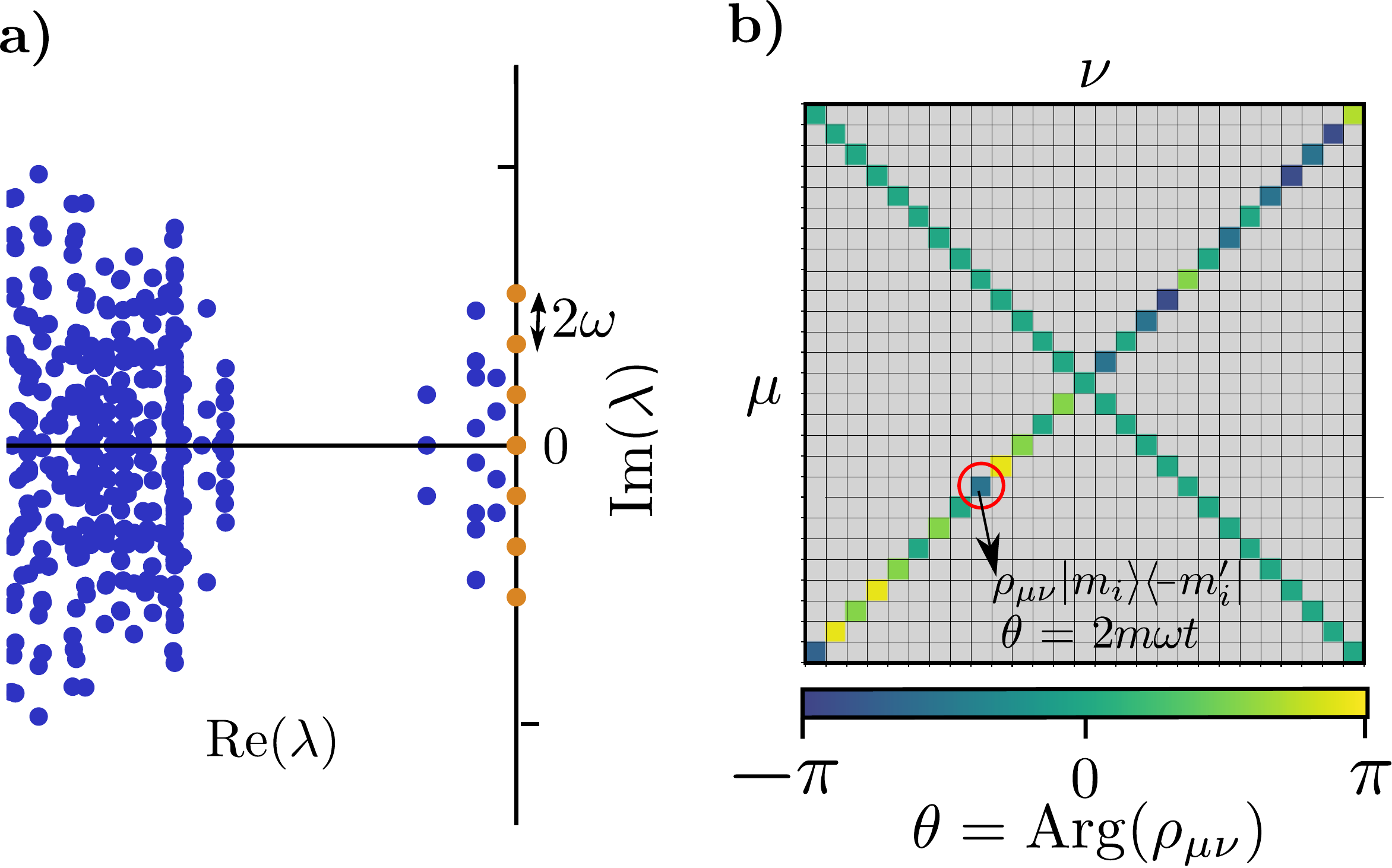}
\caption{(a) Eigenvalues $\{\lambda\}$ close to the real axis for the Liouvillian superoperator from the master Equation in Eq. (\ref{Eq:Master Equation Specific}). Parameters are $N = 3, \ \omega_{j} = 1.0J \ \forall j, \ \Delta = 0.5J, \ \gamma = 2.0J$. Eigenvalues with ${\rm Re}(\lambda)=0$ are marked in orange, all others are marked in blue. (b) Structure of the density matrix at time $tJ = 100.0$ under the map in Eq. (\ref{Eq:Master Equation Specific}) with the same parameters as in a). Initial state is a random product state. The colour indicates the phase of each complex element $\rho_{\mu \nu}$, the grey colour indicates ${\rm Abs}(\rho_{\mu \nu}) = 0$ and so the phase is not plotted. The indices $\mu$ and $\nu$ run over the basis vectors of the Hilbert space in lexicographic order when they are converted to ternary strings with $\uparrow = 2$, $0 = 1$, $\downarrow = 0$. As an example when $\mu = 1$ this corresponds to the basis vector $\ket{\uparrow \uparrow \uparrow} = \ket{222}$ and when $\mu = 27$: $\ket{\downarrow \downarrow \downarrow} = \ket{000}$. Example element (ringed in red) has a phase which evolves in time as $\theta = 2m\omega t$ where $m$ is the magnetisation of the corresponding basis vector $\ket{m_{i}}\bra{-m_{i}'}$.}
\label{Fig:Spectrum}
\end{figure}

\par The long-time dynamics of $\mathcal{L}$ is not, however, solely governed by this steady state. We identify (see SM) multiple strong dynamical symmetry operators of the form
\begin{equation}
A_{m} = \sum_{i}\ket{m_{i}}\bra{-m_{i}'}, \quad m \neq 0, m= -N, ..., N,
\label{Eq:Specific Imaginary Modes}
\end{equation}
which each satisfy Eq. (\ref{Eq:Conditions})
\begin{equation}
[H, A_{m}] = 2m\omega A_{m}, \quad [L_{j}, A_{m}] = [L_{j}^{\dagger}, A_{m}] = 0, \quad \forall j.
\label{Eq: SpinConditions}
\end{equation}
Following this we can determine, see Eq. (\ref{Eq: Imaginary Modes}), the imaginary eigenmodes of the Liouvillian through the action of these operators on the steady state.
Explicitly, we have,
\begin{equation}
\rho^{m}_{1,0} \propto A_{m}\rho_{ss} \propto A_{m}, \qquad \mathcal{L}\rho^{m}_{1,0} = -2im\omega,
\label{Eq:SpinSpectrum}
\end{equation}
which is a non-trivial result as the steady-state $\rho_{ss}$ is inherently singular. Further application of $A_{m}$ is redundant as $A_{m}A_{m} = 0$ and thus, each $A_{m}$ generates a unique imaginary eigenmode $\rho^{m}_{1,0}$ via left-multiplication of the steady-state. Crucially, however, the eigenspectrum is still equidistant as the eigenvalues of the different modes form a ladder with a spacing of $2 \omega$. Hence, the structure of the long-time eigenspace is analogous to a system with a single strong dynamical symmetry operator.
\par The steady state and imaginary eigenmodes in Eqs. (\ref{Eq:Steady State}) and (\ref{Eq:Specific Imaginary Modes}) form a complete basis for the long-time dynamics of Eq. (\ref{Eq:Master Equation Specific}) and so, similarly to Eq. (\ref{Eq: DMEvolution}), the density matrix can be expressed as a superposition of these modes in the limit $t \rightarrow \infty$.  The imaginary modes describe coherences between sectors of opposite magnetisation, their excitement will ensure the system reaches a limit cycle in the long-time limit.  Moreover, the density matrix is completely translationally invariant; as described in Sec. 2B the dephasing and interactions have washed out any geometry in the system, which now has no characteristic length-scale. This invariance can be seen in the off-diagonal coherences described in Eq. (\ref{Eq: Imaginary Modes}), which occur at all length-scales of the chain and are completely uniform with respect to distance. 

\par In Fig. \ref{Fig:Spectrum} we visualise the analytical results in Eqs. (\ref{Eq:Steady State}), (\ref{Eq:Specific Imaginary Modes}) and (\ref{Eq:SpinSpectrum}). We present a plot of the eigenspectrum of $\mathcal{L}$ [Fig. \ref{Fig:Spectrum}(a)], the formation of these imaginary eigenmodes is clear and their spacing is set by the value of $\omega$. We also show the structure of the density matrix in the long-time limit of Eq. (\ref{Eq:Master Equation Specific}) [Fig. \ref{Fig:Spectrum}(b)]. The system is in a superposition of the steady state in Eq. (\ref{Eq:Steady State}) and the imaginary modes in Eq. (\ref{Eq:Specific Imaginary Modes}), hence it only has elements along the diagonal and anti-diagonal in the configuration basis. The magnitude of each of these matrix elements is constant in time. The phase of the elements along the anti-diagonal is well-defined and evolves in time at a frequency $f = 2m\omega$ for the corresponding matrix element $\ket{m_{i}}\bra{-m'_{i}}$.  

\par We can explicitly prove that this coherent density matrix structure leads to observable synchronisation in the long-time limit of the system. Specifically, consider the operator $X = \prod_{j \in B}(S^{x}_{j})^{2}$, where $B = \{a,b,c, ...\}$ is a set of $M$ sites containing no duplicates. The operator is formed from quadratic, local operators which measure fluctuations in the magnetisation. The quadratic form is necessary in order to be able to measure the coherences between the basis states $\ket{\downarrow}$ and $\ket{\uparrow}$ - any operator formed solely from linear local operators will relax to stationarity. We prove (see SM) that in the limit $t \rightarrow \infty$
\begin{equation}
\langle X \rangle = D_{0}{\rm Tr}(\rho_{ss}X) + \sum_{m = 1}^{M}D_{m}\cos(2m\omega t){\rm Tr}(\rho^{m}_{1,0}X),
\label{Eq:X}
\end{equation}
where the real coefficients $D_{m}$ are those associated with the overlap between the initial state $\rho(0)$ and either the imaginary eigenmodes or the steady state. The trace overlap ${\rm Tr}(\rho^{m}_{1,0}X)$ only depends on the cardinality of $B$, not the specific sites within the set; a direct consequence of the complete translational invariance of the modes spanning the kernel. Consequently $\langle X \rangle$ is also independent of the specific choice of sites over which we measure the correlator $X$, only the number of sites matters. Thus we see that the long-time dynamics will perfectly synchronise the spin-1s to clean, coherent limit-cycles - regardless of the specific values of the initial state or the Liouvillian parameters.
\par Moreover, the modes in Eq. (\ref{Eq:Specific Imaginary Modes}) are entangled and cannot be written as a superposition of separable states - which we explicitly show in the following numerics. We also demonstrate that the reduced correlator $\langle X \rangle - \prod_{j \in B}\langle X_{j} \rangle$ is distance-invariant and non-zero. Hence, we consider the synchronisation observed in Eq. (\ref{Eq:X}) to be inherently quantum - underpinned by long-range correlations, which are a result of entanglement between the bodies in the system.
\par In Eq. (\ref{Eq:Hamiltonian}), the $S^{z}_{i}S^{z}_{i+1}$ (ZZ) term is an interaction term and hence the parameter $\Delta$ sets the strength of the interactions\footnote{We refer to the $S_i^+ S_{i+1}^{-}+ {\rm h.c}$ terms as hopping terms - they only lower and raise spin on neighbouring sites and so, in the $z$-basis, do not represent a true interaction term.} and plays a critical role in the formation of synchronisation. Specifically, the ZZ interaction ensures that only translationally invariant strong dynamical symmetries and steady states are present. Therefore, the asymptotic time-dependent density matrix also possesses this symmetry which, in turn, implies perfect synchronization (see Section 2B). Provided $\Delta \neq 0$ this will always be case, with the explicit value of $\Delta$ only effecting the time-scale on which synchronisation occurs. When $\Delta = 0$ there are additional steady states and strong dynamical symmetries of the Liouvillian (see SM for an example) which are not translationally invariant and interfere with the symmetry of the known solutions described earlier, disrupting the synchronicity of the system. 

\par All of these results are valid for any arbitrary length chain of $N$ spin-1s under the Liouvillian in Eq. (\ref{Eq:Master Equation Specific}). In the subsequent numerics we focus on a small series of spin-1s, i.e. $N = 3$ or $N = 4$. This is because as the system sizes increases, for generic initial states (product states for example), the diagonal correlations become increasingly dominant in the long-time limit compared to the off-diagonal coherences ($|D_{0}| \gg |D_{m \neq 0}|$), reducing the amplitude of the synchronisation measures in the system (this amplitude will, however, remain finite for any finite-size system). Consequently, by focussing on a small chain, we can readily resolve the features of quantum synchronisation and directly witness our analytical calculations by solving the master equation in Eq. (\ref{Eq:Master Equation Specific}) through numerical exponentiation of the Liouvillian superoperator $\mathcal{L}$. Later in the text we will present our second example where synchronisation is induced via our mechanism and is observable even in the thermodynamic limit. The combination of these two examples emphasizes how this symmetry-induced synchronisation occurs in systems of varying size and structure.

\begin{figure}[t!]
\includegraphics[width = \columnwidth]{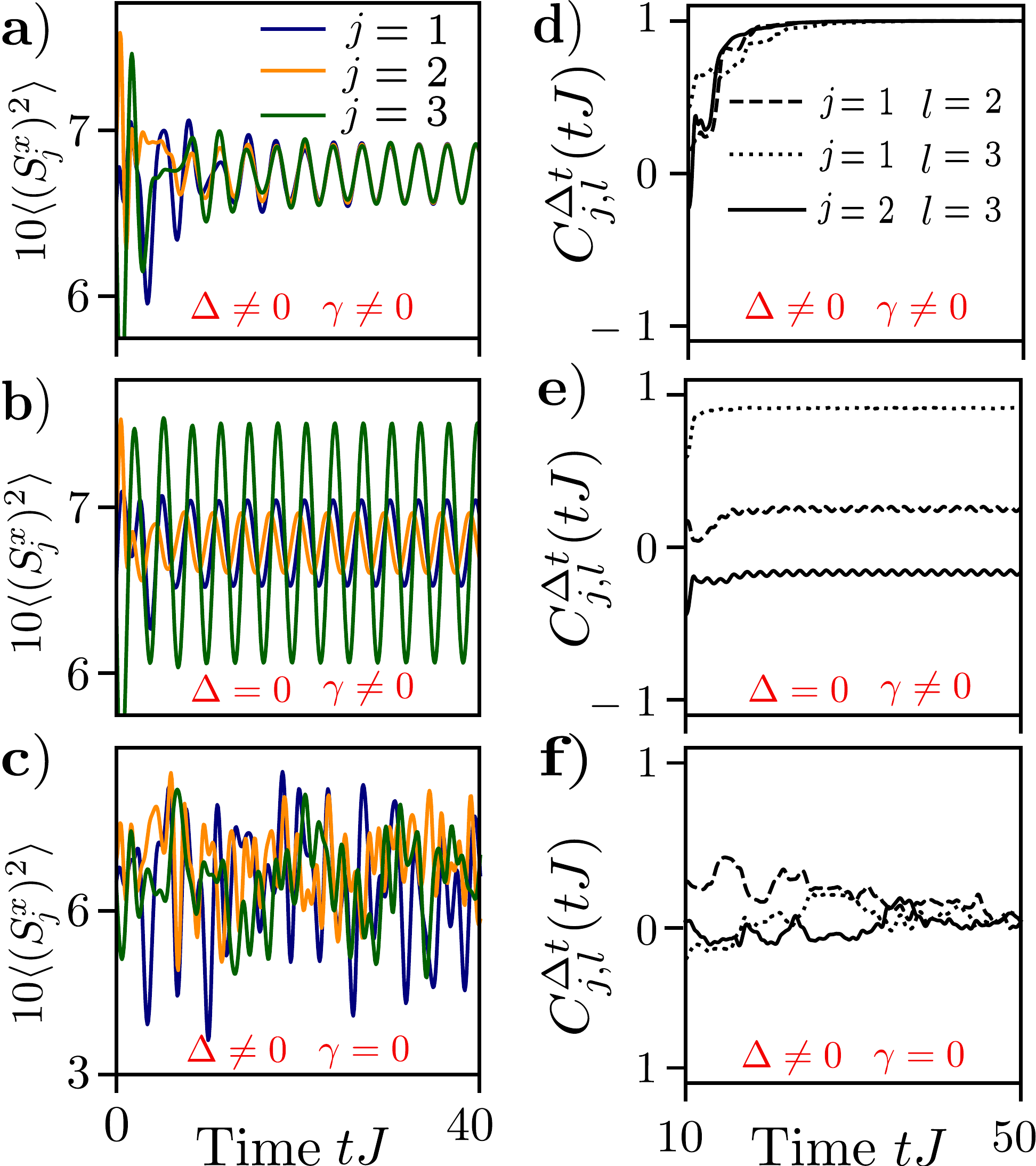}
\caption{(a-c) Dynamics of $\langle (S^{x}_{j})^{2} \rangle$ for a quench from a completely random product state under the map in Eq. (\ref{Eq:Master Equation Specific}) with $N = 3, \ \omega_{j} = 1.0J \ \forall j$. a) $\Delta = 0.5J, \ \gamma = 1.0J$, b) $\Delta = 0, \ \gamma = 2.0J$ c) $\Delta = 0.5J, \ \gamma = 0$. (d - f) Pearson time-correlation coefficient for each possible pair of functions from the respective plots in (a-c). The time-averaged window is a rolling window with width $\Delta t = 10.0tJ$ centred at time $tJ$.}
\label{Fig:Dynamics}
\end{figure}

\par For the following results, we start in a specified initial state and then time-evolve it under the Liouvillian in Eq. (\ref{Eq:Master Equation Specific}) measuring various time-dependent quantities in order to observe the formation of synchronisation.  As a first synchronisation measure for the local observables in our model we consider the time-dependent Pearson-correlation factor \cite{QSynchMeasures, QSynch6}. It can be used to measure the correlation over time for two functions $f, g$ defined on a domain $[t, t + \Delta t]$
\begin{equation}
C^{\Delta t}_{f, g}(t) = \frac{\int_{t}^{t + \Delta t}(f - \bar{f})(g - \bar{g})dt}{\sqrt{\int_{t}^{t+\Delta t}(f - \bar{f})^{2}dt   \int_{t}^{t+\Delta t}(g - \bar{g})^{2}dt}},
\label{Eq: Pearson}
\end{equation} 
with the function average $\bar{f} = \frac{1}{\Delta t}\int_{t}^{t + \Delta t}fdt$. This correlation factor is maximal (minimal), $1$ ($-1$) when the two signals $f$ and $g$ are perfectly synchronised (anti-synchronised), and $0$ when they display no correlations.  

\begin{figure}[t!]
\includegraphics[width = \columnwidth]{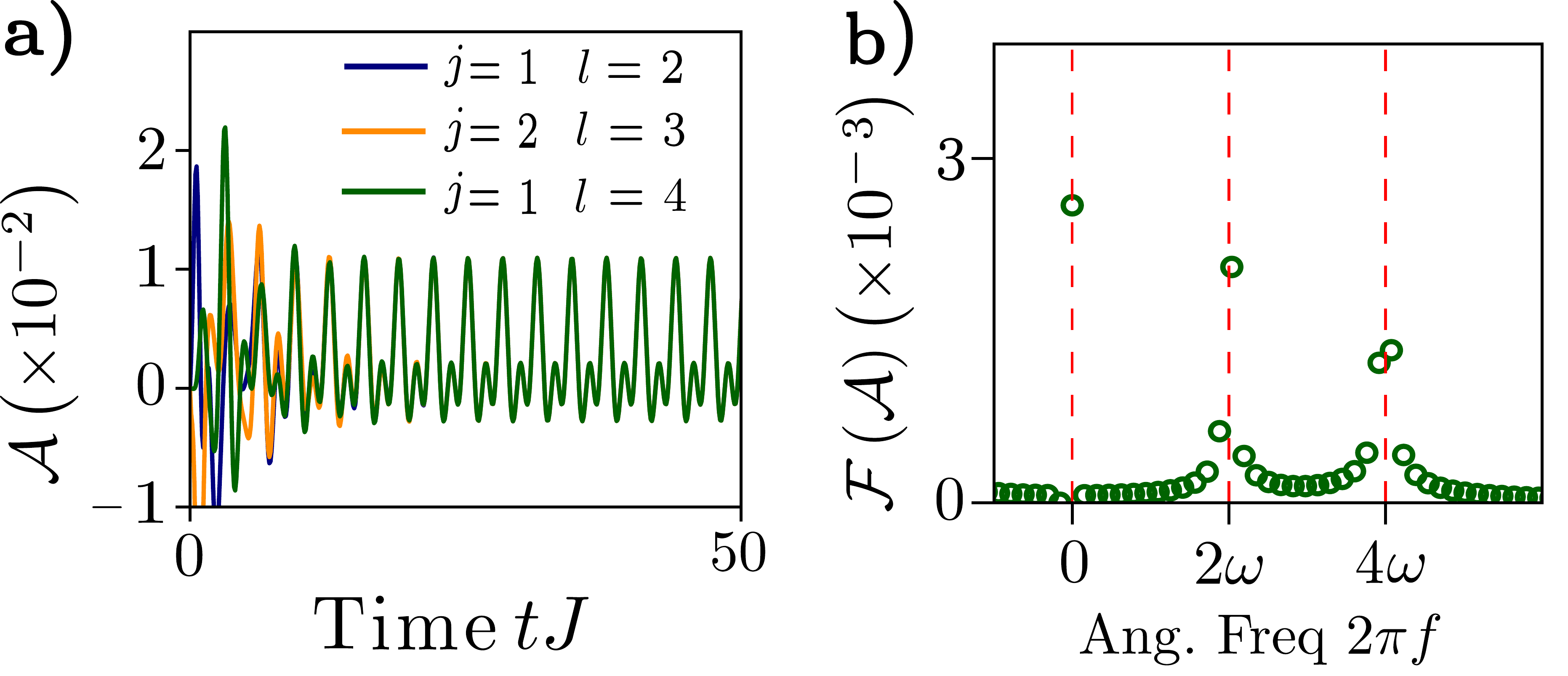}
\caption{Dynamics of $\mathcal{A} = \langle (S^{x}_{j})^{2}(S^{x}_{l})^{2} \rangle - \langle (S^{x}_{j})^{2} \rangle \langle (S^{x}_{l})^{2} \rangle$ for a quench from a random state with total $x$-magnetisation $\langle S^{x} \rangle = 0$ under the map in Eq. (\ref{Eq:Master Equation Specific}) with $N = 4, \ \omega_{j} = 1.0J \ \forall j, \ \Delta = 0.5J, \ \gamma = 2.0J$ b) Amplitude of the Fourier transform $\mathcal{F}(\mathcal{A})$ versus angular frequency in the long-time limit. The red-dashed lines indicate the expected angular frequency response based on Eq. (\ref{Eq:X}). Only a selection of the possible values of $j$ and $l$ are shown for brevity.}
\label{Fig:TwoPointDynamics}
\end{figure}

\par In Fig. \ref{Fig:Dynamics} we set $f = \langle (S^{x}_{j})^{2} \rangle$ and $g = \langle (S^{x}_{l})^{2} \rangle$ in order to measure the synchronisation over time between two of the spin-1s $j$ and $l$ - we start from a completely random product state. We also include the individual functions $\langle (S^{x}_{j})^{2} \rangle$ over time for each spin. In agreement with  Eq. (\ref{Eq:X}), when both the environment and interactions are present ($\gamma,\ \Delta \neq 0$) the dynamics causes the spins to synchronise perfectly [Figs. \ref{Fig:Dynamics}(a) and (d)] to the same frequency and phase, despite being initialised with random phases. The frequency of the oscillations is directly  determined by the equidistant spacing of the imaginary eigenvalues. For comparison [Fig. \ref{Fig:Dynamics}(b)] we show the case when the environment is present but there are no interactions ($\gamma \neq 0, \Delta = 0$). The presence of dissipation causes the system to converge to the expected clean coherent limit cycles \cite{DarkHamiltonians} but the limit cycles for each spin are out of phase. Despite the fact the system still has a strong dynamical symmetry, the absence of interactions prevents synchronisation as the kernel of the Liouvillian still has a memory of the initial geometry of the system. We also show the closed case when $\gamma = 0$ [Figs. \ref{Fig:Dynamics}(c) and (f)], the dynamics are completely chaotic and unsynchronised due to the multitude of incommensurate frequencies in the eigenvalues of the Hamiltonian.

\par Whilst the plots in Fig. \ref{Fig:Dynamics} demonstrate that the spins are able to perfecly lock phases they do not capture the collective origin of this synchronisation. In this vein, in Fig. \ref{Fig:TwoPointDynamics}, we plot the reduced correlator $\mathcal{A} = \langle (S^{x}_{j})^{2}(S^{x}_{l})^{2} \rangle - \langle (S^{x}_{j})^{2} \rangle \langle (S^{x}_{l})^{2} \rangle$, which is non-zero and identical for any choice of spins $j$ and $l$. The observed oscillations contain several frequencies [Fig. \ref{Fig:TwoPointDynamics}(b)] due to the excitement of multiple imaginary modes in Eq. (\ref{Eq:Specific Imaginary Modes}). These imaginary modes all contain coherences between different spins, the synchronisation observed in Fig. \ref{Fig:Dynamics} is dependent on the existence of these inter-spin coherences and Fig. \ref{Fig:TwoPointDynamics} shows that they give rise to perfect, distance-invariant correlations throughout the system.

\begin{figure*}[t!]
\includegraphics[width = \textwidth]{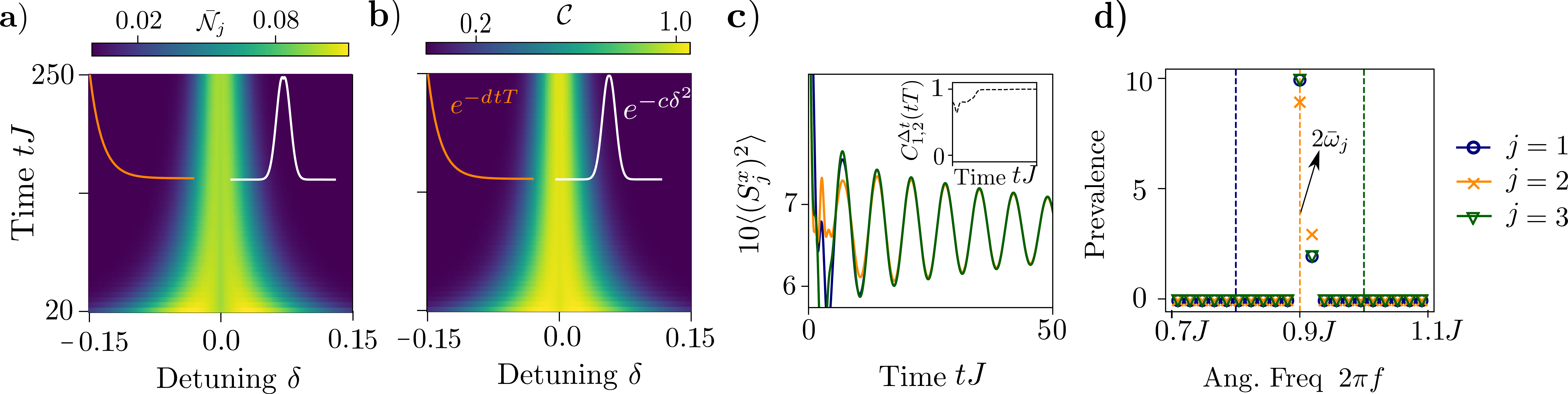}
\caption{a-b) Plot of synchronisation witnesses vs time and detuning for the master Equation in Eq. (\ref{Eq:Master Equation Specific}) with $N = 3$. The system is initialised in the state $\ket{\psi(0)} = \ket{\rightarrow 0 0}$, where $\ket{\rightarrow} = \frac{1}{\sqrt{2}}(\ket{\uparrow} + \ket{\downarrow})$, and evolved in time with parameters $\{\omega_{1}, \omega_{2}, \omega_{3}\} = \{1.0 - \delta, 1.0, 1.0 + \delta\}J, \ \gamma = 2.0J, \Delta = 0.5J$. Insets, top-left) Cross-section (orange) of these measures versus time at detuning $\delta = -0.075$. top-right) Cross-section (white) of these measures versus detuning at time $tJ = 250.0$. The parameters $d$ and $c$ are constants used to parametrise the cross-sections (see SM).  a) Synchronisation is measured by the average negativity, see Eq. (\ref{Eq: NegativityCoherence}), for each site. b) Synchronisation is measured as the total magnitude of the off-diagonal coherences,  see Eq. (\ref{Eq: NegativityCoherence}). (c) Example dynamics of $\langle (S^{x}_{j})^{2} \rangle$ for the same system except with specific natural frequencies $\{\omega_{1}, \omega_{2}, \omega_{3}\} = \{0.4, 0.45, 0.5\}J$ and dephasing $\gamma = 1.0J$. Inset) Pearson Coefficient, see Eq. (\ref{Eq: Pearson}), for the two functions $\langle (S^{x}_{1})^{2} \rangle$ and $\langle (S^{x}_{2})^{2} \rangle$ from time $tJ = 5$ to $tJ = 20$. d) Prevalence of angular frequencies, extracted from the distribution of angular frequencies created using the the time-periods between successive turning points for the oscillations in c) but up to $tJ = 100.0$. Dashed lines indicate the expected delta function in the prevalence, based on each spin's natural frequency. The central line is twice the average of the natural frequencies $\bar{\omega}_{j}$.}
\label{Fig:Tongue}
\end{figure*}

\par So far we have considered the `homogeneous' case $\omega_{j} = \omega \ \forall j$. The spins share the same natural frequency and we have shown how, under dephasing and interactions, their phases will align perfectly. In order to discuss synchronisation in full we now set the frequencies of the spins to be mismatched. In this case, the imaginary modes in Eq. (\ref{Eq: Imaginary Modes}) are no longer exact eigenvectors of the Liouvillian and, in the long-time limit, the system will decay to an ensemble which is diagonal in the configuration basis - where no synchronisation can occur. However, as was shown in Eq. (\ref{Eq:ModLiouvillian}), for sufficiently small values of $\epsilon$, the system is only slightly perturbed from a `dynamical symmetry' regime defined by the spins having a common frequency which is the average of their natural frequencies. Furthermore due to the translational symmetry of the imaginary eigenmodes, we know the system is, to first order, completely robust to this kind of perturbation. Hence, it is interesting to observe whether the spins are able to synchronise to the dynamical symmetry regime on an intermediate time-scale and, if so, how long the system takes to desynchronise and reach a diagonal ensemble.
\par As measures to track this, and to highlight the quantum nature of the synchronisation in Figs. \ref{Fig:Dynamics} and \ref{Fig:TwoPointDynamics}, we introduce two common witnesses for quantum synchronisation: the negativity $\mathcal{N}$ \cite{Negativity} and off-diagonal coherences $\mathcal{C}$ \cite{QSynch3}
\begin{align}
\mathcal{N}_{j}(\rho) &= \frac{||\rho^{T_{j}}|| - 1}{2}, \notag \\  \mathcal{C} &= \sum_{i \neq j}|\rho_{ij}|,
\label{Eq: NegativityCoherence}
\end{align}
with $T_{j}$ indicating the partial transpose with respect to site $j$ and $||X|| = {\rm Tr}\sqrt{X^{\dagger}X}$ denoting the trace norm of an operator. 
The negativity can be seen as a measure of the degree to which spin $j$ is entangled with the rest of the system whilst the coherence quantifier $\mathcal{C}$ describes the total magnitude of the off-diagonal elements in the density matrix. When the frequencies are matched the system is synchronised, and due to the off-diagonal, entangled nature of the modes in Eq. (\ref{Eq: Imaginary Modes}) quantities such as these will remain finite indefinitely.
\par In Fig. \ref{Fig:Tongue} we show how these synchronisation witnesses evolve in time when the system is perturbed from the dynamical symetry regime. We use the detuning strength $\delta$ to characterise the range of the natural frequencies. The explicit distribution of natural frequencies is not important, the key parameter is its width and in the SM we obtain similar results when the natural frequencies are drawn from a uniform random distribution. Initially, the system is in a product state where $\mathcal{N} = 0$, the transient dynamics then causes the formation of entanglement and anti-diagonal coherences which decay away at a rate set by $\delta$. We show how this entanglement forms [Figs. \ref{Fig:Tongue}(c-d)]: despite having mismatched frequencies and phases the spins lock to an intermediate limit cycle with identical phase and frequency - which is twice the average of the natural frequencies $\bar{\omega}_{j}$ (due to the factor of $2$ in Eq. (\ref{Eq: SpinConditions})).  The life-time of this cycle is large and as $\delta \rightarrow 0$ diverges to infinity, evidenced by the tongue-like behaviour seen in Figs. \ref{Fig:Tongue}(a-b). These figures show how the corresponding measures act as strong witnessess to the synchronisation in the system - emphasizing its quantum nature.

\begin{figure}[t!]
\includegraphics[width = \columnwidth]{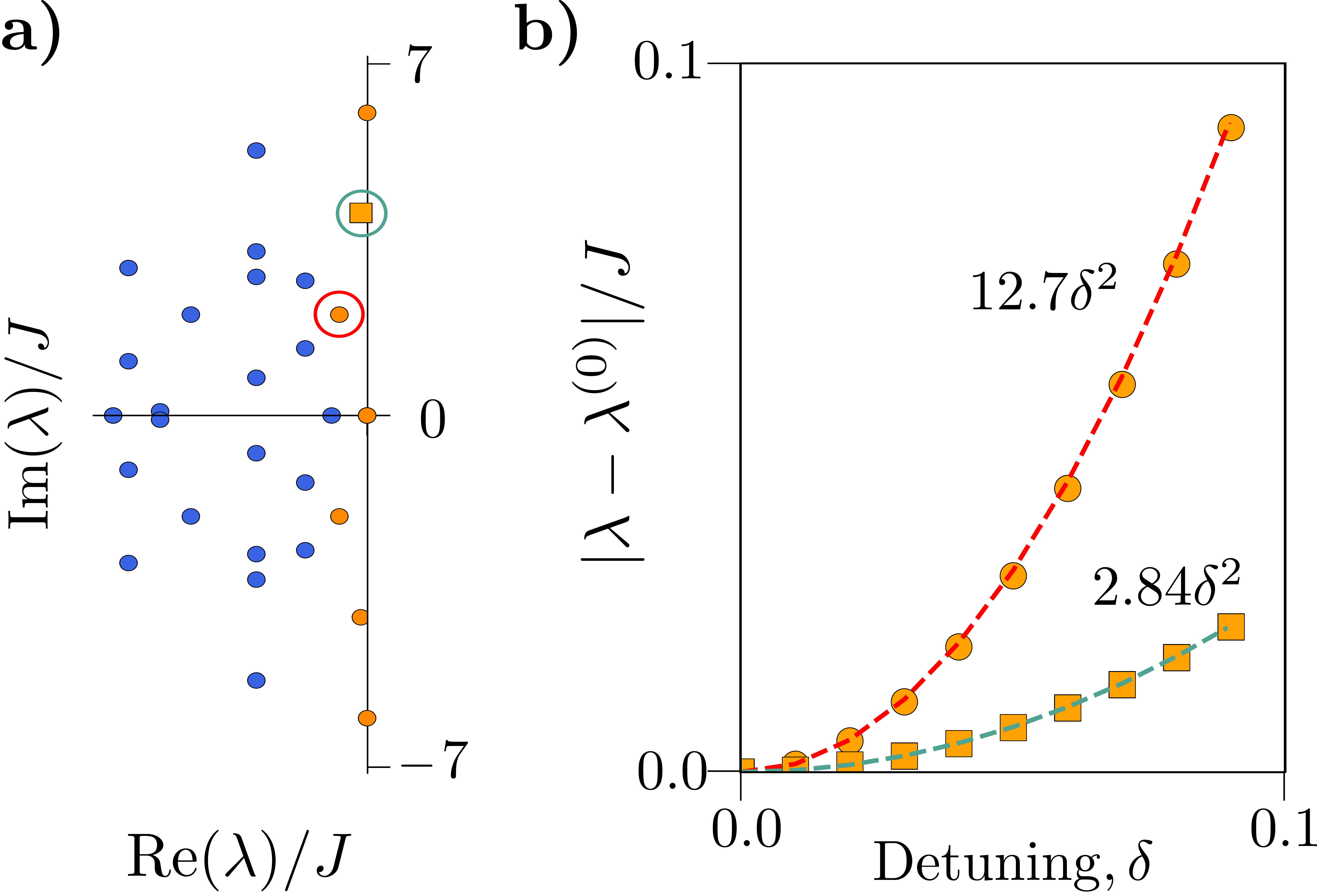}
\caption{a) Spectrum of eigenvalues close to the real axis for the Liouvillian in Eq. (\ref{Eq:Master Equation Specific}) with $\Delta = 0.5J$, $\gamma = 2.0J$, $\{\omega_{1}, \omega_{2}, \omega_{3}\}$ = $\{1.0 - \delta, 1.0, 1.0 + \delta\}J$ and $\delta = 0.07$. Orange coloured eigenvalues are those which lie on the imaginary axis when $\delta = 0$, they are shifted from the axis due to the finite value of $\delta$. (b) Scaling of the distance shifted $|\lambda - \lambda^{(0)}|$, in the complex plane, as a function of $\delta$, where $\lambda$ is the new eigenvalue and $\lambda^{(0)}$ is the original imaginary eigenvalue when $\delta = 0$. Curves are for the two circled eigenvalues in a), which shows the specific example $\delta = 0.07$. Dashed curves are quadratic fits.}
\label{Fig:Scaling}
\end{figure}

The imaginary eigenmodes in Eq. (\ref{Eq: Imaginary Modes}) are translationally-invariant and hence the robust, synchronised behaviour observed [Fig. \ref{Fig:Tongue}] is the result of a second-order response to the detuning. The cross-sections included in Fig. \ref{Fig:Tongue}b are evidence of this. At a given time the coherences are well-approximated by a gaussian profile (see SM) as a function of the detuning. Meanwhile at a given detuning the coherences decay away exponentially as a function of time, the decay rate $d$ is proportional to the square of the detuning (see SM for numerical evidence of this). Furthermore, we explicitly show this non-linear scaling in Fig. \ref{Fig:Scaling}. We calculate the shift in the imaginary eigenvalues, $|\lambda - \lambda^{(0)}|$ (see Sec. (\ref{Sec: Perturbations away from the Dynamical Symmetry Regime})) from their original value $\lambda^{(0)}$ at $\delta = 0$ as a function of the detuning $\delta$. There is no noticeable shift to first order in $\delta$ - the fitted curve is proportional to $\delta^{2}$. Notably, the highest imaginary eigenmode $\rho^{N}_{1,0} = \ket{\uparrow \uparrow ... }\bra{\downarrow \downarrow ...}$ is always unshifted, and remains imaginary regardless of the distribution of natural frequencies.

\subsection{Many-body synchronisation in the Hubbard model}
As our second example, we take the $1$D $N$-site Hubbard model \cite{HubbardModel} in, potentially, disordered magnetic and chemical fields. We focus on 1D lattices for numerical tractability, nonetheless it should be emphasized that these results are solely based on symmetry and thus can be observed in any bi-partite $d$-dimensional realisation of the Hubbard model. The Hamiltonian reads
\begin{align}
H = -\tau\sum_{\langle jl \rangle, \sigma}(c^{\dagger}_{\sigma, j}c_{\sigma, l} + {\rm h.c}) + \ U\sum_{j}n_{\uparrow, j}n_{\downarrow, j} \ \notag \\ + \frac{1}{2}\sum_{j}\omega_{j}(n_{\uparrow, j} - n_{\downarrow, j}) + \sum_{j}\mu_{j}(n_{\uparrow, j} + n_{\downarrow, j}),
\label{HubbardHam}
\end{align} 
where $c_{\sigma, j}^{\dagger}$ and its adjoint are the usual creation and annihilation operators for a fermion of spin $\sigma \in \{\uparrow, \downarrow\}$ on site $j$. Additionally, $n_{\sigma, j}$ is the number operator for a particle of spin $\sigma$ on site $j$ and $\tau$, $U$, $\omega_{j}$ and $\mu_{j}$ play the role of kinetic, interaction, magnetic and chemical energy scales respectively.
\par We then couple the system to a bath which induces spin-agnostic dephasing on each site. Hence, the system's time evolution can be described by the Lindblad Equation: 
\begin{align}
\frac{\partial \rho}{\partial t} = \mathcal{L}\rho = -i[H, \rho] + \gamma \sum_{j = 1}^{N}n_{j}\rho n_{j} - \frac{1}{2}\{(n_{j})^{2}, \rho\}, \ \notag \\ \qquad n_{j} = n_{j, \uparrow} + n_{j, \downarrow}.
\label{Eq:Master Equation Specific2}
\end{align}

\begin{figure}[t]
\includegraphics[width = \columnwidth]{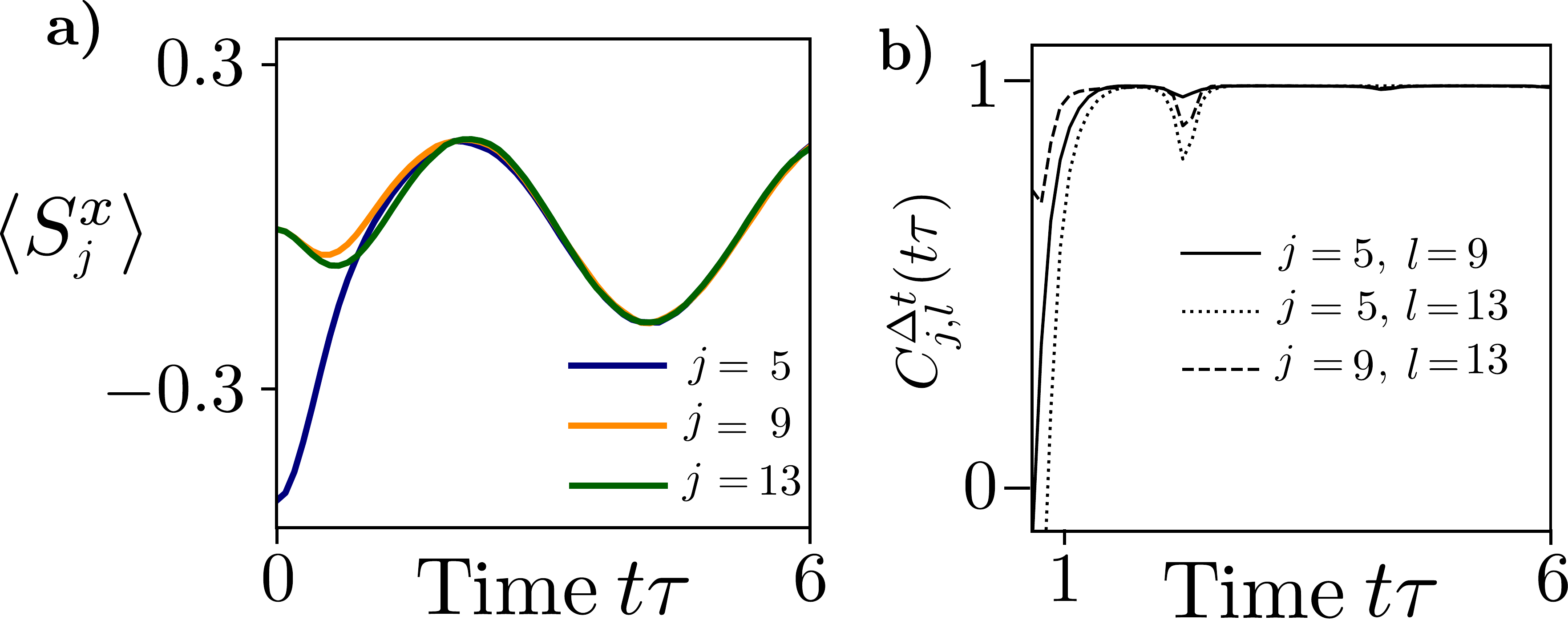}
\caption{a) Dynamics of $\langle S^{x}_{j} \rangle$ for a quench of the $N = 15$ site charge-dephased Hubbard model described by the master Equation in Eq. (\ref{Eq:Master Equation Specific2}). Parameters are $\gamma = 2.5\tau, \ U = 1.0\tau, \ \omega_{j} = 1.5\tau \ \forall j, \ \mu_{j} \in {\rm Rand}[0.0, 0.2]\tau$ where ${\rm Rand}[0.0, 0.2]\tau$ is a uniformly-drawn random number on the specified interval.  The initial state is $\ket{\psi(0)} = \otimes_{j = 1}^{5} \ket{\chi}$ where $\ket{\chi} = \ket{\leftarrow \downarrow \uparrow}$ and $\rightarrow, \leftarrow, \uparrow$ and $\downarrow$ correspond to each site being polarised in the positive $x$-direction, negative $x$-direction, positive $z$ and negative $z$-direction respectively. b) Pearson time-correlation coefficient for each possible pair of functions from a). The time-averaged window is a rolling window with width $\Delta t = 0.5t\tau$ centred at time $t\tau$.}
\label{Fig:DynamicsHubbard}
\end{figure}

\begin{figure*}[t]
\includegraphics[width = \textwidth]{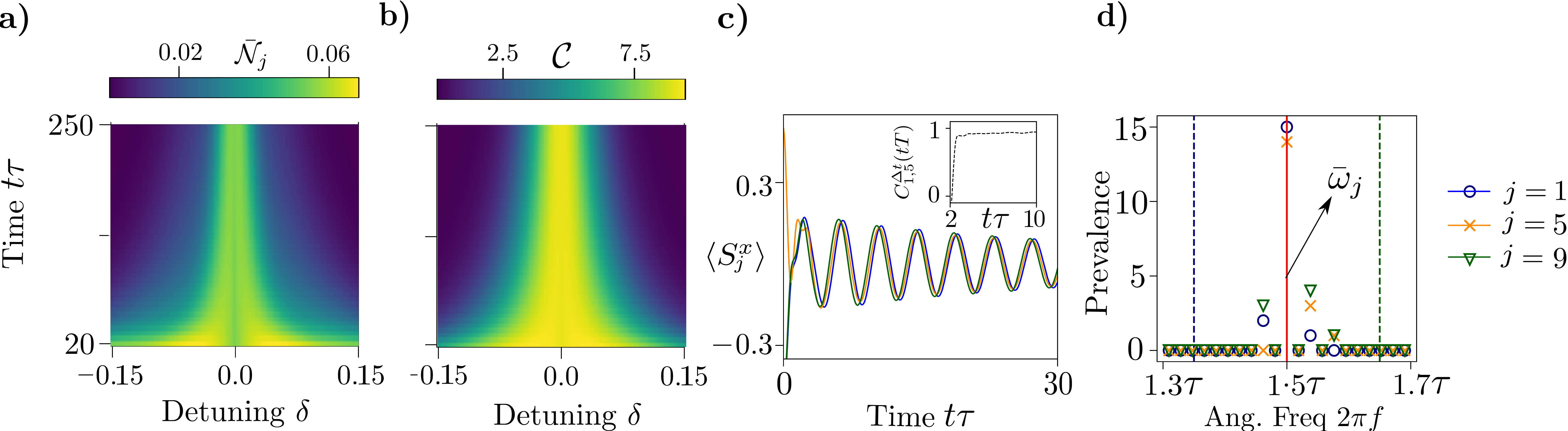}
\caption{a-b) Plot of synchronisation witnesses vs time and detuning for the charge-dephased Hubbard model with $N = 5 $ sites. The system is initialised in the state $\ket{\psi(0)} = \ket{\rightarrow \uparrow \rightarrow \uparrow \rightarrow}$, which has no double occupancies, where $\rightarrow, \leftarrow, \uparrow$ and $\downarrow$ correspond to each site being polarised in the positive $x$, negative $x$, positive $z$ and negative $z$ directions respectively. The system is then evolved in time with parameters $\gamma = 2.0\tau, U = 0.5\tau$ and  $\mu_{j} \in {\rm Rand}[0,0.2]\tau$ where ${\rm Rand}[0,0.2]$ is a uniformly-drawn random number on the specified interval. The magnetic frequencies are uniformly distributed $\{\omega_{1}, \omega_{2}, \omega_{3}, \omega_{4}, \omega_{5}\} = \{1.0 - \delta, 1.0 - \delta/2, 1.0, 1.0 + \delta/2, 1.0 + \delta\}\tau$. a) Synchronisation is measured by the average negativity for each site. b) Synchronisation is measured as the total magnitude of the off-diagonal coherences. (c) Example dynamics of $\langle S^{x}_{j} \rangle$ for $N=9$. The natural frequencies of the Hamiltonian are: $\{\omega_{1}, \omega_{2}, ...\} = \{1.35, 1.3875, ...,  1.65\}\tau$ and other parameters are $\gamma = 0.5\tau, \ U = 1.0\tau$, $\mu_{j} \in [0.0, 0.2]\tau$. The starting state is $\ket{\psi(0)} = \otimes_{j = 1}^{3}\ket{\chi}$ where $\ket{\chi} = \ket{\rightarrow \leftarrow \rightarrow}$. Inset) Pearson Coefficient for the two functions $\langle S^{x}_{1} \rangle$ and $\langle S^{x}_{5} \rangle$ over time with a rolling window of $\Delta t = 2.0t\tau$. d) Prevalence of different angular frequencies, extracted from the distribution of angular frequencies created using the the time-periods between successive turning points for the oscillations in c) but up to $t\tau = 90.0$. Blue and green dashed lines indicate the expected delta function in the prevalence based on the natural frequency of spins $1$ and $9$. Red solid line indicates the average of all the natural frequencies $\bar{\omega}_{j} $. Only a subset of sites are represented in plots c-d) for clarity.}
\label{Fig:TongueHubbard}
\end{figure*}

\par The case $\omega_{j} = \omega \ \forall j$ of this model was originally studied in Ref. \cite{DarkHamiltonians}, where the existence of a strong dynamical symmetry was shown to ensure long-time non-stationary dynamics in this strongly-correlated system. A possible experimental realisation of the system is also described in this reference. In this section we show how, further to this, perfect synchronisation is induced in the  long-time dynamics of this model: alongside the strong dynamical symmetry there is an inter-site coupling and homogeneous local dephasing which ensure the long-time dynamics is completely cooperative and translationally invariant. Moreover, later in the section we break the homogeneity of the magnetic field
and  demonstrate the robustness of this synchronisation to perturbations away from the dynamical symmetry regime.
\par Firstly, when $\omega_{j} = \omega \ \forall j$, the magnetic field in Eq. (\ref{HubbardHam}) breaks the spin ${\rm SU}(2)$ symmetry of the model: $[H, S^{\pm}] = \omega S^{\pm}$, $\ [L_{j}, S^{\pm}] = 0 \ \forall j$, where $S^{+} = \sum_{j}c_{j, \uparrow}^{\dagger}c_{j, \downarrow}$ is the global magnetic raising operator. Hence, there is a single strong dynamical symmetry operator which can be used to form the set of equidistant imaginary eigenmodes
\begin{equation}
\rho_{nm} \propto (S^{+})^{n}\rho_{ss}(S^{-})^{m}, \ \mathcal{L}\rho_{nm} = i\omega (m-n)\rho_{nm}.
\label{Eq: HubbardModes}
\end{equation} 
\par These modes cause the existence of a persistent limit cycles in the magnetisation (in the $x$ and $y$ directions) of the system. For example, by defining the operator $X = \prod_{j \in B}S^{x}_{j}$ (where $B=\{a,b,c ..\}$ is a set of $M$ sites containing no duplicates) we can use Eq. (\ref{Eq: Evolution}) to prove (see SM):
\begin{equation}
\lim_{t \rightarrow \infty}\langle X \rangle (t) = \sum_{\substack{i = 0}}^{\lfloor M/2 \rfloor}D_{i}\cos\left( \left( 2i+d \right) \omega t \right), \ d = M \ {\rm mod} \ 2,
\label{Eq:XHub}
\end{equation}
where the coefficients $D_{i}$ are set by the initial state of the system. Due to the inter-site coupling and local dephasing the imaginary modes in Eq. (\ref{Eq: HubbardModes}) are completely translationally symmetric (see Ref. \cite{DarkHamiltonians} for the explicit form of the steady state) and thus this observable is depent only on the cardinality $M$ of the set $B$, not the specific sites within the set. 

\par As a result, even in the presence of disorder in the chemical potential, the system displays perfectly synchronised magnetic oscillations in the long-time limit. This will occur for a wide range of specific parameters and initial states of the system, the only requirements are that the appropriate coefficients, $D_{i}$, are finite and the hopping amplitude $\tau$ (which couples the different sites together) is non-zero. Moreover, in the thermodynamic limit, initial states which have $\lim_{N \rightarrow \infty}\langle S^{x} \rangle/N \neq 0$ will ensure the long-time synchronised oscillations in $\langle S^{x}_{j} \rangle$ or $\langle S^{x}_{jS}S^{x}_{j+1} \rangle$  have a finite amplitude \cite{DarkHamiltonians}. Similarly to the previous example these oscillations are underpinned by long-range correlations in the system which arise due to the entangled nature of the long-time density matrix. We demonstrate these features in the following numerics, showing synchronisation in a fully many-body quantum system. 

\par In order to increase the system size accessible to our numerical calculations, we have used a ‘quantum trajectories’ approach \cite{Trajectories} to perform a stochastic unravelling of Eq. (\ref{Eq:Master Equation Specific2}) and simulate the dynamics at the level of an ensemble of pure wavefunctions. Furthermore in Fig. \ref{Fig:DynamicsHubbard}, as the simulation is only on a short time-scale, we were able to use the time-evolving block decimation \cite{TEBD} algorithm on a Matrix Product State \cite{MPS} decomposition of the trajectory wavefunctions, further increasing the available system size. These simulations were performed with the aid of the Tensor Network Theory library \cite{TNT}.

\par In Fig. \ref{Fig:DynamicsHubbard} we demonstrate the synchronicity which results from the eigenmodes in Eq. (\ref{Eq: HubbardModes}). We initialise the system in a product state and, after quenching under the master Equation in Eq. (\ref{Eq:Master Equation Specific2}) observe how the $x$-magnetisation on each site synchronises perfectly, oscillating at the anticipated frequency. The Pearson coefficient for the magnetisation on any two sites saturates to $1$ in the long-time limit [Fig. \ref{Fig:DynamicsHubbard}(b)], with the dip at $t \tau \approx 2$ being a transient effect which occurs at the first turning point in the magnetisation.  
\par We now perturb the system from the dynamical symmetry regime by setting the natural frequencies $\omega_{j}$ to be inhomogeneous, here we draw them from an evenly spaced distribution, i.e. $\omega_{j+1} - \omega_{j} = {\rm const}$ and $\omega_{N} - \omega_{1} = 2\delta$. Again, as for the spin-1 case, we choose this distribution for simplicity, our observations are independent of the explicit distribution - the key parameter is its width $\delta$. We initialise the system in a specified state and time-evolve under the Liouvillian in Eq. (\ref{Eq:Master Equation Specific2}). In Fig. \ref{Fig:TongueHubbard} we show how, similarly to the previous spin-1 example, the system is still attracted to the synchronised state, in both phase and frequency, on an intermediate time-scale. There is a significant band of detunings where the system stays in this long-lived synchronisation phase (Figs. \ref{Fig:TongueHubbard}a and b) and the spin on each site locks to the same phase and frequency (which is set by the average frequency of the individual sites, see Fig. \ref{Fig:TongueHubbard}d). Remarkably, this harmonized response is occuring even in the presence of both magnetic and chemical disorder - emphasizing the robustness of a symmetry-based approach to observing quantum synchronisation. The imaginary eigenmodes in Eq. (\ref{Eq: HubbardModes}) are translationally-invariant and hence $\lambda^{(1)}$ evaluates to $0$ (see Sec. \ref{Sec: Perturbations away from the Dynamical Symmetry Regime}). As with the previous example, this robust, synchronised behaviour is a result of a second-order response to the detuning.

\section{Conclusion}
We have provided condtions which, when satisfied, guarantee synchronisation in a generic qopen quatnum system. We have then shown how, using a combination of analytics and numerics, the interplay between interactions, local dephasing and a strong dynamical symmetry can satisfy these conditions and facilitate the combination of entanglement and perfect phase synchronisation between the individual constituents of the system. This is a direct result of the formation, in the long-time limit, of a well-defined phase relationship in the off-diagonal coherences of the density matrix, at all length-scales of the system. Furthermore, when perturbed from the dynamical symmetry regime these systems exhibit a second-order response which results in both phase and frequency locking throughout the system.
\par These observations orginate at the level of the symmetries of the system. Thus, we believe, this work marks an important step in understanding how fully-quantum synchronisation can originate in a wide range of generic physical systems - as opposed to in a single delicately controlled setup. We anticipate further examples of complex quantum networks where symmetry can guide the individual nodes into an entangled, fully synchronised state. 
\par We highlight the potential role such a harmonised response can play in developing quantum technologies such as atomic clocks and other metrological instruments - which rely on quantum-enhanced synchronicity and cooperative behaviour in order to outperform their classical counterparts \cite{Qtech1, Qtech2}.
\par We also consider it pertinent to explore the role of dynamical symmetries in synchronising closed, strongly-correlated systems - where dissipation is absent. Recently, it was shown how the presence of a quasi-local dynamical symmetry can prevent stationarity and guarantee oscillatory dynamics in the XXZ model \cite{Marko}. 
\par Finally we note that strong dynamical symmetries are also defined outside of this approximation \cite{DarkHamiltonians} and so we anticipate our results to be extendible beyond the Markov regime. This will be particularly important for understanding whether our results can be observed for non-local dissipation: strongly interacting subsystems are often not described by local master equations \cite{DeChiara2018}. 
\newline 
\section{Acknowlegdments}
We would like to thank J. Mur-Petit and J. Coulthard for useful discussions. This work has been supported by EPSRC grants No. EP/P009565/1 and EP/K038311/1 and is partially funded by the European Research Council under the European Union’s
Seventh Framework Programme (FP7/2007-2013)/ERC Grant Agreement No. 319286 Q-MAC. In
carrying out this work we acknowledge the use of the University of Oxford Advanced Research Computing (ARC) facility  http://dx.doi.org/10.5281/zenodo.22558, the QuTiP Python toolbox for simulating open quantum systems http://qutip.org \cite{qutip} and the Tensor Network Theory library \cite{TNT} for performing the TEBD \cite{TEBD} algorithm which produced the data in Fig. \ref{Fig:DynamicsHubbard}.

\pagebreak 
\pagebreak
\clearpage

\section*{Supplemental material to ``Quantum Synchronisation Enabled by Dynamical Symmetries and Dissipation"}

\section{Perturbation Theory on a Frequency Detuned Liouvillian}
We consider a general Liouvillian formed from the Lindblad equation $\mathcal{L}\rho = -i[H, \rho] + D[\rho]$. The Hamiltonian contains, amongst other terms, an inhomogeneous field $\sum_{j}\omega_{j}f_{j}$ where $f_{j}$ is some local field operator. We can split the Hamiltonian into two terms, a homogeneous part: $\sum_{j}\bar{\omega}_{j} f_{j}$ and an inhomogeneous part: $\sum_{j}\delta_{j} f_{j}$ where $\omega_{j} = \bar{\omega}_{j} + \delta_{j}$ and $\bar{\omega}_{j}$ is the average natural frequency. We then, correspondingly, split the Liouvillian into a perturbed and an unperturbed part, scaling by $1/\bar{\omega}_{j}$:
\begin{align}
\mathcal{L} = \mathcal{L}^{(0)} + \epsilon \mathcal{L}^{(1)}, \qquad \epsilon = \frac{\bar{\delta}_{j}}{\bar{\omega}_{j}}, \notag \\ \mathcal{L}^{(0)} = -\frac{i}{\bar{\omega}_{j}}[H - \sum_{j}\delta_{j} f_{j}, \bigcdot] + \frac{1}{\bar{\omega}_{j}}D[\bigcdot], \notag \\ \mathcal{L}^{(1)} = -i \left[ \sum_{j}\frac{\delta_{j}}{\bar{\delta}_{j}} f_{j}, \bigcdot \right],
\end{align}
with $\bar{\delta}_{j}$ is the average over the set of detunings $\{\delta_{j}\}$.
\par We work in superket and superbra form and assume that the unperturbed Liouvillian $\mathcal{L}^{(0)}$ contains a series of imaginary eigenvectors and values $\mathcal{L}^{(0)}|\rho^{(0)}_{i} \rangle \rangle = \lambda^{(0)}_{i} |\rho_{i} \rangle \rangle$, $ \ {\rm Re}(\lambda^{(0)}_{i}) = 0$, indexed by $i$. We let $\langle \langle \sigma^{(0)}_{i}|$ denote the corresponding left-eigenvectors $\langle \langle \sigma^{(0)}_{i}|\mathcal{L}^{(0)} = \lambda^{(0)}_{i}\langle \langle \sigma^{(0)}_{i}|$. For small $\epsilon << 1$ we expand the eigenvectors and values of the new Liouvillian $\mathcal{L}$ as a peturbative power series on the previous, i.e.
\begin{align}
|\rho_{i} \rangle \rangle &= |\rho_{i}^{(0)} \rangle \rangle + \epsilon |\rho_{i}^{(1)} \rangle \rangle + \epsilon^{2}|\rho_{i}^{(2)} \rangle \rangle + \hdots, \notag \\
\langle \langle \sigma_{i} | &= \langle \langle \sigma_{i}^{(0)}| + \epsilon \langle \langle \sigma_{i}^{(1)}| + \epsilon^{2} \langle \langle \sigma_{i}^{(2)} | + \hdots, \\ \notag
\lambda_{i} &= \lambda_{i}^{(0)} + \epsilon \lambda_{i}^{(1)} + \epsilon^{2}\lambda_{i}^{(2)} + \hdots.
\label{Eq:Expansions}
\end{align}
We also know that the orthonormality condition $\langle \langle \sigma_{i} | \rho_{j} \rangle \rangle = {\rm Tr}(\sigma^{\dagger}_{i}\rho_{j}) = \delta_{i,j}$ must hold $\forall \epsilon$ - where we have defined $\sigma_{i}$ and $\rho_{j}$ as the matrix forms of the corresponding superket and superbras. Using this condition, to 0th and 1st order, we have
\begin{align}
{\rm Tr} \left( \left(\sigma_{i}^{(0)}\right)^{\dagger}\rho_{j}^{(0)} \right) = {\rm Tr} \left( \left(\sigma_{i}^{(1)}\right)^{\dagger}\rho_{j}^{(0)} + \left(\sigma_{i}^{(0)}\right)^{\dagger}\rho_{j}^{(1)} \right) \notag \\ = \delta_{ij}.
\end{align}
We now simplify the known expression $\langle \langle \sigma_{i}| \mathcal{L}| \rho_{i} \rangle \rangle = \lambda_{i} \langle \langle \sigma_{i} |\rho_{i} \rangle \rangle$ by subsitituting the expansions in Eq. (\ref{Eq:Expansions}) and converting to matrix form. As a result we find the first order correction to the imaginary eigenvalue
\begin{align}
\lambda^{(1)}_{i} = {\rm Tr} \left( \left(\sigma_{i}^{(0)} \right)^{\dagger}\mathcal{L}^{(1)}\rho_{i}^{(0)} \right).
\end{align}
For imaginary eigenmodes formed from a strong dynamical symmetry \cite{DarkHamiltonians} it can be proved that the left and right eigenmodes are the same, i.e. $\langle \langle \sigma_{i}^{(0)} | = \langle \langle \rho_{i}^{(0)} |$ and so 
\begin{align}
\lambda^{(1)}_{i} &= {\rm Tr} \left( \left(\rho_{i}^{(0)} \right)^{\dagger}\mathcal{L}^{(1)}\rho_{i}^{(0)} \right) \notag \\ &= -i\sum_{j}\frac{\delta_{j}}{\bar{\delta}_{j}} {\rm Tr} \left( \left(\rho_{i}^{(0)} \right)^{\dagger}f_{j} \rho_{i}^{(0)}- \left(\rho_{i}^{(0)} \right)^{\dagger}\rho_{i}^{(0)}f_{j}  \right).
\label{Eq:Trace}
\end{align}
If the mode $\rho_{i}^{(0)}$ is translationally invariant (i.e. it is unchanged under a permutation of any pair of sites) then we notice that the trace in Eq. (\ref{Eq:Trace}) is independent of $j$. Using the fact $\sum_{j} \delta_{j} = 0$ we then have $\lambda^{(1)}_{i} = 0$. Hence, for translationally-invariant imaginary eigenmodes formed from a strong dynamical symmetry we find that the system exhibits a non-linear response to perturbations in the homogeneity of the natural frequencies. This underpins the strong-synchronised response of the two systems considered in the main text.

\section{Imaginary Modes and Steady States of a Spin 1 Chain}
Here we prove the existence of certain imaginary modes and steady states of a dephased XXZ spin-1 chain of length $N$. In the main text, we consider the Lindblad equation
\begin{align}
\frac{\partial \rho}{\partial t} = \mathcal{L}\rho &= -i[H, \rho] + \gamma \sum_{j = 1}^{N}(S^{z}_{j})^{2}\rho (S^{z}_{j})^{2} - &\frac{1}{2}\{(S^{z}_{j})^{4}, \rho\} \notag \\ &= -i[H, \rho] + D[\rho]
\label{Eq:Master Equation Specific}
\end{align}
with the Hamiltonian $H$
\begin{align}
H = \omega \sum_{j = 1}^{N} S^{z}_{j} + \sum_{j = 1}^{N-1}J\big(S^{+}_{j}S^{-}_{j+1} + S^{-}_{j}S^{+}_{j+1}\big) + \Delta S^{z}_{j}S^{z}_{j+1}.
\label{Eq:Hamiltonian}
\end{align}
We start by proving that any state $\rho = \sum_{i =1}^{G_{m}}\ket{m_{i}}\bra{m_{i}}$, where $\ket{m_{i}}$ is one of the $G_{m}$ eigenvectors satisfying $S^{z}\ket{m_{i}} = m\ket{m_{i}}$, is a steady state: $\mathcal{L}\rho = 0$. Firstly we substitute $\rho_{ss}$ into Eq. (\ref{Eq:Master Equation Specific}) where it is easy to show that $[S^{z}_{j}, \ket{m_{i}}\bra{m_{i}}] = [S^{z}_{j}S^{z}_{j+1}, \ket{m_{i}}\bra{m_{i}}] = D[\ket{m_{i}}\bra{m_{i}}] = 0, \ \forall m, i, j$. Hence, it remains to show
\begin{align}
J\sum_{j = 1}^{N-1}[S^{+}_{j}S^{-}_{j+1} + S^{-}_{j}S^{+}_{j+1}, \sum_{i =1}^{G_{m}}\ket{m_{i}}\bra{m_{i}}] = 0.
\label{Eq:SSEq1}
\end{align}
This can be done by considering the two spin-1s on the $j$ and $j +1$ positions for a given $\ket{m_{i}}$. Then, we have that $S^{+}_{j}S^{-}_{j+1}\ket{m_{i}}$ is a non-zero vector only if the two spin-1s are in one of the configurations $\ket{0\uparrow}, \ket{00}, \ket{\downarrow \uparrow}, \ket{\downarrow 0}$. Because for these configurations swapping spins $j+1$ and $j$ doesn't change the magnetisation $m$, we can always find the term $\ket{m_{i'}}\bra{m_{i'}}$ in the steady state where $\ket{m_{i'}}$ is just $\ket{m_{i}}$ with spins $j$ and $j+1$ swapped. Equation (\ref{Eq:SSEq1}) then follows from the fact $S^{+}_{j}S^{-}_{j+1}\ket{m_{i}}\bra{m_{i}} - \ket{m_{i'}}\bra{m_{i'}}S^{+}_{j}S^{-}_{j+1} = 0$, i.e. for every term we can find a corresponding term to cancel it with.
\par We can also show that $\mathcal{L}\sum_{i = 1}^{G_{m}}\ket{m_{i}}\bra{-m_{i}'} = 2i m \omega \sum_{i = 1}^{G_{m}}\ket{m_{i}}\bra{-m_{i}'}$, where $\bra{-m_{i}'}$ is the `spin-flipped' bra for $\ket{m_{i}}$ (i.e. if $\ket{2_{1}} = \ket{0 \uparrow \uparrow}$ then $\ket{-2_{1}'} = \ket{0 \downarrow \downarrow}$). Firstly, it is clear that $[\sum_{j}S^{z}_{j}, \ket{m_{i}}\bra{-m_{i}'}] = 2m\ket{m_{i}}\bra{-m_{i}'}$.  Secondly,  we also have $[S^{z}_{j}S^{z}_{j+1}, \ket{m_{i}}\bra{-m_{i}'}] = D[\ket{m_{i}}\bra{-m_{i}'}] = 0 \ \forall m, i, j$. Finally, by a very similar agument (the term with the $j$ and $j + 1$ spins swapped can always be found in $\sum_{i = 1}^{G_{m}}\ket{m_{i}}\bra{-m_{i}'}$) to the previous paragraph we find that $\sum_{i = 1}^{G_{m}}\ket{m_{i}}\bra{-m_{i}'}$ satisfies Eq. (\ref{Eq:SSEq1}) in the same manner as $\sum_{i =1}^{G_{m}}\ket{m_{i}}\bra{m_{i}}$. 
\par Hence we have another possible steady state: $\mathcal{L}\sum_{i = 1}^{G_{0}}\ket{0_{i}}\bra{0_{i}'} = 0$ and so we can write the full steady state as
\begin{align}
\rho_{ss} = \sum_{m = -N}^{N}\bigg(\sum_{i  = 1}^{G_{m}}\lambda_{m}\ket{m_{i}}\bra{m_{i}}\bigg) + \lambda_{0}'\sum_{i = 1}^{G_{0}}\ket{0_{i}}\bra{0'_{i}},
\label{Eq:Steady State}
\end{align}
which is $2N+2$ fold degenerate. The coefficients $\{\lambda_{m}\}$ and $\lambda_{0}$ must satisfy
\begin{equation}
\lambda_{0}' + \sum_{m = -N}^{N}\lambda_{m}\sum_{s = 0}^{N}{N \choose s}{N - s \choose (N- s + m)/2} = 1,
\end{equation}
in order for ${\rm Tr}(\rho_{ss}) = 1$. The terms in the second summation are skipped if $(N- s + m)/2$ is not an integer.
\par Furthermore, the imaginary eigenmodes $\rho^{m}_{1,0} = \sum_{i = 1}^{G_{m}}\ket{m_{i}}\bra{-m_{i}'} \ m \neq 0, m= -N, ..., N$, which we proved satisfy $\mathcal{L}\rho^{m}_{1,0} = 2i n \omega \rho^{m}_{1,0}$, originate as a series of strong dynamical symmetries \cite{DarkHamiltonians} of the model because:
\begin{align}
[H, \rho^{m}_{1,0}] = 2\omega m \rho^{m}_{1,0}, \quad [L_{k}, \rho^{m}_{1,0}] = [L_{k}^{\dagger}, \rho^{m}_{1,0}] = 0 \quad \forall k, m.
\end{align}
I.e. for this system the strong dynamical symmetry operators are the imaginary modes because they return themselves upon application to the steady state $\rho^{m}_{1,0}\rho_{ss} \propto \rho^{m}_{1,0}$ (the steady state is singular). Further application of the strong dynamical symmetry operators is redundant as $\rho^{m}_{1,0}\rho^{m}_{1,0} = 0$. For the case when $N > 2$ and $\Delta \neq 0$ numerical calculations show that these $2N+2$ steady states and $2N$ imaginary modes completely span the kernel of $\mathcal{L}$ and thus form a complete description of the system's dynamics in the limit $t \rightarrow \infty$.

\par \textit{Noninteracting Solutions} - The steady states and strong dynamical symmetries described above provide a complete basis for the long-time density matrix of the system when $\Delta \neq 0$. When $\Delta = 0$ these solutions are still valid (the derivations of the previous section are true $\forall \Delta$), however there exist additional solutions which are not translationally invariant and therefore disrupt the synchronicity of the system. This is because the system is no longer interacting due to the absence of the $S^{z}_{i}S^{z}_{i+1}$ term (in the $z$-basis the $S^{+}_{i}S^{-}_{i+1}$ terms only describe the exchange of excitations through the lattice and do not constitute interaction terms).
\par We now provide an example of one of these translationally invariant solutions. Specifically one can define the operator
\begin{equation}
B = \sum_{i=1}^{N}(-1)^{i}\ket{\uparrow \uparrow ... \uparrow}\bra{\uparrow \uparrow ... \uparrow}S_{i}^{+},
\end{equation}
and easily prove that
\begin{align}
&[H(\Delta = 0), B] = 2B, \quad [H(\Delta \neq0), B] \neq \lambda B \quad \lambda \in \mathbb{Re}, \notag \\ &[L_{j}, B] = 0 \ \forall j,
\label{Eq: Symmetry Breaking}
\end{align}
where $H(\Delta)$ is the Hamiltonian in Eq. (\ref{Eq:Hamiltonian}) as a function of $\Delta$. The first $2$ relations in Eq. (\ref{Eq: Symmetry Breaking}) are because the interaction term in the Hamiltonian $\sum_{j}S^{z}_{i}S^{z}_{i+1}$ does not commute with $B$ whilst the hopping term does. Hence $B$ is only a valid dynamical symmetry operator when $\Delta = 0$. The operator $B$ is clearly not translationally invariant (even and odd sites are distinct) and so this interferes with the perfect translational invariance of the solutions derived earlier. Moreover, there are additional solutions which break the translational invariance further. When these solutions are excited by the initial state then the synchronicity of the system is disrupted (see Fig. 3 of the main text).

\begin{figure*}[t]
\includegraphics[width = \textwidth]{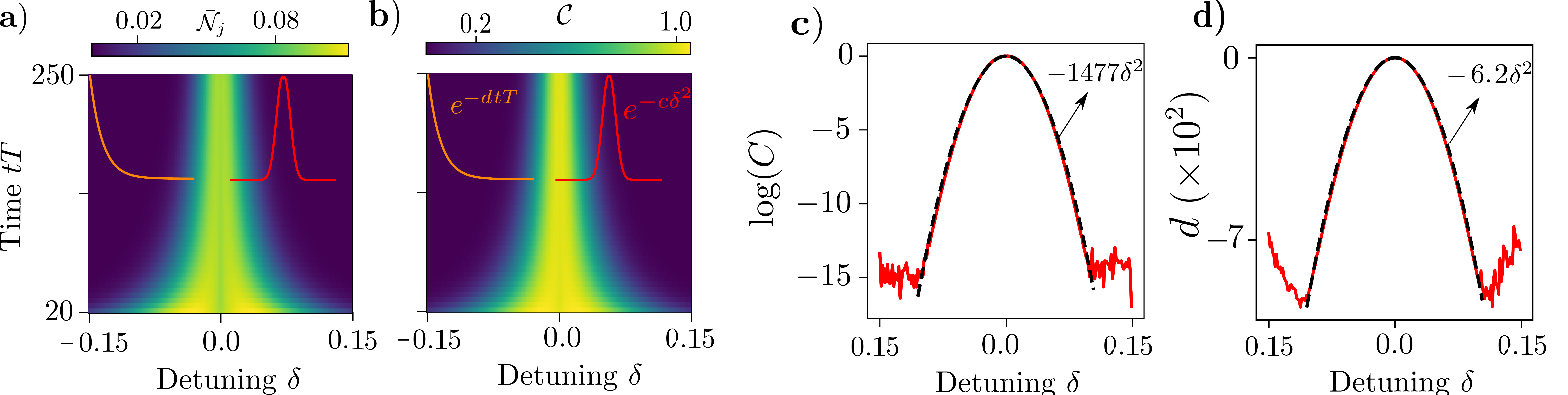}
\caption{a-b) Reproduced from main text, plot of synchronisation witnesses vs time and detuning for the dephased spin-1 chain described by Eq. (\ref{Eq:Master Equation Specific}) and  with $N = 3$. The system is initialised in the state $\ket{\psi(0)} = \ket{\rightarrow 0 0}$, where $\ket{\rightarrow} = \frac{1}{\sqrt{2}}(\ket{\uparrow} + \ket{\downarrow})$, and evolved in time with parameters $\{\omega_{1}, \omega_{2}, \omega_{3}\} = \{1.0 - \delta, 1.0, 1.0 + \delta\}J, \ \gamma = 2.0J, \Delta = 0.5J$. Insets, top-left) Cross-section (orange) of these measures versus time at detuning $\delta = -0.075$. top-right) Cross-section (orange) of these measures versus detuning at time $tJ = 250.0$. The parameters $d$ and $c$ are used to parametrise the cross-sections.  a) Synchronisation is measured by the average negativity for each site. b) Synchronisation is measured as the total magnitude of the off-diagonal coherences. c) Natural logarithm of the off-diagonal coherences $\mathcal{C}$ versus the detuning, taken from the inset in the top right of b). Dotted line is a quadratic fit over $\delta \in [-0.1, 0.1]$ d) Decay coefficient $d$ (fitted to the exponential decay of $\mathcal{C}$ versus time $tT$) versus detuning. Dotted line is a quadratic fit over $\delta \in [-0.1, 0.1]$.}
\label{Fig:S1}
\end{figure*}

\section{Long-Time Dynamics of the Spin 1 Chain}
As the imaginary modes $\rho^{m}_{1,0}$ contain coherences between the states $\ket{\uparrow}$ and $\ket{\downarrow}$ then they will only affect the dynamics of quadratic observables such as $(S^{x}_{j})^{2}$ and $(S^{y}_{j})^{2}$. We can always write the long-time density matrix as
\begin{align}
\lim_{t \rightarrow \infty}\rho(t) = C_{0} \rho_{ss} + \left(\left(\sum_{m = 1}^{N}e^{2i\omega m t} C_{m} \rho^{m}_{1,0}\right) + {\rm h.c.}\right),
\end{align}
where $C_{m}$ are a series of real coefficients (to ensure hermicity) associated with the overlap between the initial state $\rho(t = 0)$ and either the steady state $\rho_{ss}$ or the imaginary modes $\rho^{m}_{1,0}$. We also have $(\rho^{m}_{1,0})^{\dagger} = \rho^{-m}_{1,0}$. We consider the expectation value of the operator  $(S^{x}_{j})^{2} = (1/4)(S^{+}_{j} + S^{-}_{j})^{2}$. As the imaginary modes for which $|m| \geq 2$ must contain at least two flipped spins between the states $\ket{m_{i}}$ and $\bra{-m_{i}}$ then we immediately have ${\rm Tr}(\rho^{m}_{1,0}(S^{x}_{j})^{2}) = 0, \ |m| \geq 2 \ \forall j$. Hence, we get:
\begin{align}
\lim_{t \rightarrow \infty} \langle (S^{x}_{j})^{2} \rangle (t) &= C_{0}{\rm Tr}(\rho_{ss}(S^{x}_{j})^{2}) \notag \\ &+ 2C_{1}\cos(2\omega t){\rm Tr}(\rho_{1}(S^{x}_{j})^{2}),
\label{Eq: SxExpec}
\end{align}
where we have used the fact ${\rm Tr}(\rho_{-1}(S^{x}_{j})^{2}) = {\rm Tr}(\rho_{1}(S^{x}_{j})^{2})$. Equation (\ref{Eq: SxExpec}) proves the formation of clean, single frequency oscillations in the associated observable. Furthermore, the modes $\rho_{ss}$ and $\rho^{m}_{1,0}$ are all translationally invariant and so the oscillations are identical for all spins: ensuring perfect phase synchronisation.
\par In order to observe the excitement of higher order modes we must measure higher order correlators. Specifically consider the operator 
\begin{align}
X = \prod_{j \in \{a,b,c ..\}}(S^{x}_{j})^{2}, \quad \qquad |\{a, b, c, ..., \}| = M,
\end{align}
where the set of $M$ sites $\{a, b, c, ..., \}$ contains no duplicates. Because we now have ${\rm Tr}(\rho^{m}_{1,0}X) \neq 0, \ m \leq M$ and ${\rm Tr}(\rho^{m}_{1,0}X) = {\rm Tr}(\rho^{-m}_{1,0}X)$ then we find 
\begin{align}
\langle X \rangle = C_{0}{\rm Tr}(\rho_{ss}X) + 2\sum_{m = 1}^{M}C_{m}\cos(2m\omega t){\rm Tr}(\rho^{m}_{1,0}X),
\end{align}
and see the appearance of higher order frequencies due to the excitement of higher order imaginary modes. This explains the Fourier Spectrum observed in Fig. 4b) in the main text.

\section{Parametrising the Cross-Sections of the frequency-detuned Spin-1 Chain}
In the main text we considered the response of the system when the magnetic field is inhomogeneous, i.e. the system's dynamics is modelled by Eq. (\ref{Eq:Master Equation Specific}) with the Hamiltonian now of the form
\begin{equation}
H = \sum_{j = 1}^{N} \omega_{j} S^{z}_{j} + \sum_{j = 1}^{N-1}J\big(S^{+}_{j}S^{-}_{j+1} + S^{-}_{j}S^{+}_{j+1}\big) + \Delta S^{z}_{j}S^{z}_{j+1}.
\label{Eq:HamiltonianInHom}
\end{equation}
where the $\omega_{j}$ are a series of natural frequencies associated with each spin $j$. We then considered how, for a given range of natural frequencies, synchronisation witnesses such as the negativity $\mathcal{N}_{j}(\rho) = \frac{||\rho^{T_{j}}|| - 1}{2}$ \cite{Negativity} or off-diagonal coherences $\mathcal{C} = \sum_{i \neq j}|\rho_{ij}|$ \cite{QSynch3} evolve in time. For the example in the main text we considered $N = 3$ with the natural frequencies equally spaced $\{\omega_{1}, \omega_{2}, \omega_{3}\} = \{1 - \delta, 1, 1 + \delta\}J$, which produced the maps in Fig. \ref{Fig:S1}(a-b), showing the  witnesses as a function of time and detuning.

\par We now parametrise the cross-sections in Fig. \ref{Fig:S1}(b). In Fig. \ref{Fig:S1}(c) we show how, at a given time, and for sufficiently small detunings, the off-diagonal coherences $\mathcal{C}$ are well-described by a Gaussian curve as a function of the detuning. Furthermore, in Fig. \ref{Fig:S1}(d) we calculate the decay coefficient (for the exponential decay of the off-diagonal coherences $\mathcal{C}$ versus time) versus $\delta$ and show how, for small detunings, $d \propto \delta^{2}$. In Figs. \ref{Fig:S1}(c-d) the tails of the distribution aren't captured by this parametrisation due to numerical precision (both synchronisation quantities are very close to $0$ for large detunings). This parametrisation also holds for the cross-sections of the average negativity $\bar{\mathcal{N}}_{j}$ in Fig. \ref{Fig:S1}(b).

\begin{figure*}[t]
\includegraphics[width = \textwidth]{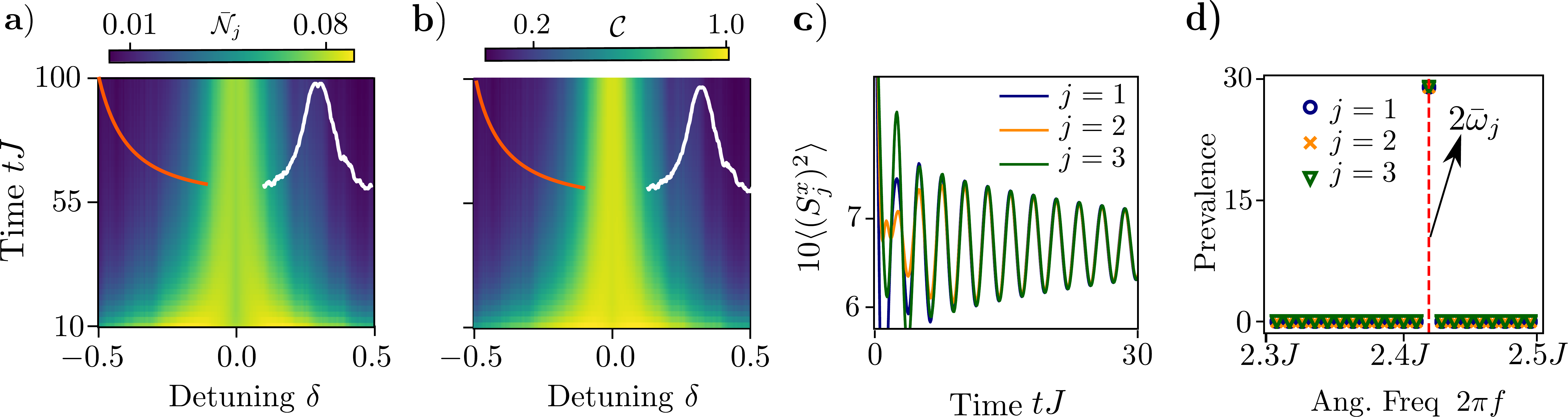}
\caption{a-b) Plot of synchronisation witnesses vs time and detuning for Eq. (\ref{Eq:Master Equation Specific}). The system is initialised in the state $\ket{\psi(0)} = \ket{\rightarrow 0 0}$ and evolved in time with parameters $\omega_{j} = (1+\epsilon_{j})J, \ \gamma = 2.0J, \Delta = 0.5J$ and random detuning $\epsilon_{j} \in [0, \delta]$ where $[0, \delta]$ is a uniform random distribution over the specified interval. The Synchronisation measures are then averaged over $100$ instances of disorder associated with the detuning. Insets) top-left, orange, shows the witness versus time at detuning $\delta = -0.25$. Top-right, white, shows the witness versus detuning at time $tJ = 55.0$. a) Synchronisation is measured by the average negativity, $\bar{\mathcal{N}}_{j}$, for each site. b) Synchronisation is measured as the total magnitude of the off-diagonal coherences $\mathcal{C}$. c) Example dynamics of $\langle (S^{x}_{j})^{2} \rangle$ for the same system except with specific natural frequencies $\{\omega_{1}, \omega_{2}, \omega_{3}\} = \{1.258, 1.210, 1.160\}J$. d) Prevalence of angular frequencies, extracted from the distribution of angular frequencies created using the the time-periods between successive turning points for the oscillations in c) but up to $tJ = 100.0$. The central line is twice the average of the spin's natural frequencies $\bar{\omega}_{j}$.}
\label{Fig:STongue}
\end{figure*}

\section{Further Plots of the frequency-detuned Spin-1 Chain}
In Figure 5 of the main text we showed how, when the natural frequencies of the spins in the chain are inhomogeneous, the system still locks to a long-lived, synchronised cycle with a frequency which is the average of their natural frequencies. This response emerges as a tongue-like profile in the witnesses $\mathcal{C}$ and $\bar{\mathcal{N}}_{j}$ as a function of detuning $\delta$ and time. In the main text, for simplicity, we considered the case where $\{\omega_{1}, \omega_{2}, \omega_{3}\} = \{1.0 - \delta, 1.0, 1.0 + \delta\}$, i.e. the natural frequencies form a uniform sequence. Here, in Fig. \ref{Fig:STongue}, we show that this distribution is arbitrary, showing how similar tongues and cross-sections emerge when the natural frequencies are drawn from a uniform random distribution of width $\delta$. The spins are able to lock to an intermediate cycle with a frequency which is twice the average of the natural frequencies $\bar{\omega}_{j}$. 

\section{Persistent Limit Cycles of the charge-dephased Hubbard Model}
Now, we turn our attention to the Hubbard model. The Lindblad equation in this case reads
\begin{align}
\frac{\partial \rho}{\partial t} = \mathcal{L}\rho &= -i[H, \rho] + \gamma \sum_{j = 1}^{N}S^{z}_{j}\rho S^{z}_{j} - &\frac{1}{2}\{S^{z}_{j}, \rho\} \notag \\  &= -i[H, \rho] + D[\rho],
\label{Eq:Master Equation Hubb}
\end{align}
with the Hamiltonian $H$
\begin{align}
H = -\tau\sum_{\langle jl \rangle, \sigma}(c^{\dagger}_{\sigma, j}c_{\sigma, l} + {\rm h.c}) + &U\sum_{j}n_{\uparrow, j}n_{\downarrow, j} + \omega \sum_{j} S_{j}^{z}, \qquad \notag \\ &S^{z}_{j} = n_{\uparrow, j} - n_{\downarrow, j}.
\label{HubbardHam}
\end{align} 
It is known \cite{DarkHamiltonians} that the imaginary eigenmodes of this Liouvillian are
\begin{align}
\rho_{nm} = (A)^{n}\rho_{ss}(A^{\dagger})^{m}, \quad \mathcal{L}\rho_{nm} = i(m-n)\omega \rho_{nm}, \quad \notag \\ A = S^{+} = \sum_{j}c_{j, \uparrow}^{\dagger}c_{j, \downarrow},
\end{align}
where $\rho_{ss}$ is a grand-canonical-like state containing the strong-symmetries of the system \cite{Prosen,DarkHamiltonians,Tindall} and $\rho_{n,m}^{\dagger} = \rho_{m,n}$. Thus in the long-time limit the state of the system can be written as
\begin{align}
\lim_{t \rightarrow \infty} \rho(t) = \sum_{\substack{n,m \\ n \geq m}}C_{n,m} e^{i(m-n)\omega t}\rho_{n,m} + {\rm h.c.},
\end{align}
where the $C_{n,m}$ are a series of real coefficients associated with the overlap between the initial state and the modes $\rho_{n,m}$.
We calculate the expectation value of the operator $S^{x}_{j}$
\begin{align}
\lim_{t \rightarrow \infty} \langle S^{x}_{j} \rangle(t) = \sum_{\substack{n,m \\ n \geq m}}C_{n,m}e^{i(m-n)\omega t}{\rm Tr}\left(S^{x}_{j} \rho_{n,m} \right) + {\rm h.c.}.
\end{align}
By expressing $S^{x}_{j}$ in terms of raising and lowering operators and using the fact that a) ${\rm Tr}\left(S^{x}_{j}\rho_{n,m}\right) = {\rm Tr}\left(S^{x}_{j}\rho_{m,n}\right)$ and b) the trace vanishes unless $|m-n|=1$ (as the operator $S^{x}_{j} \rho_{n,m}$ will have no diagonal elements in the eigenbasis of $\rho_{ss}$) we get
\begin{align}
\lim_{t \rightarrow \infty} \langle S^{x}_{j} \rangle(t) = 2\cos(\omega t)\sum_{\substack{n = 1}}C_{n,n-1}Y_{n,n-1},
\end{align}
with $Y_{n,n-1} = {\rm Tr}\left(S^{x}_{j} \rho_{n,n-1} \right) = {\rm Tr}\left(S^{x}_{j} \rho_{n-1,n} \right)$.
Hence, we see persistent oscillations in $S^{x}_{j}$, which are centred around the $x$-axis. The modes $\rho_{n,m}$ are completely translationally invariant and thus the spins on each site will synchronise to limit cycles perfectly in phase, regardless of the initial state.
We can immediately treat higher order modes through the operator
\begin{align}
X = \prod_{j \in \{a,b,..\}}S^{x}_{j}, \qquad \qquad |\{a, b, ...\}| = M,
\end{align}
where the set of $M$ sites $\{a, b, c, ..., \}$ contains no duplicates. We can calculate the expectation value of this operator by expanding it in terms of raising and lowering operators and using the fact the trace of each term is only non-vanishing if the difference between the number of raising and lowering operators is equal to $|m-n|$. Thus, we get
\begin{align}
\lim_{t \rightarrow \infty}\langle X \rangle (t) = \sum_{\substack{i = 0}}^{\lfloor M/2 \rfloor}D_{i}\cos\left( (2i+d) \omega t \right), \qquad d = M \ {\rm mod} \ 2,
\end{align}
where the $D_{i}$'s are a series of coefficients based on the initial state and the various traces between the $\rho_{n,m}$ and products of local spin-raising and lowering operators.

\end{document}